# Control of single spin-flips in a Rydberg atomic fractal

R. C. Verstraten[1]*, I. H. A. Knottnerus[2,3,4]*, Y. C. Tseng[2,3], A. Urech[2,3], T. S. do Espirito Santo[1,5], V. Zampronio[1,5,6], F. Schreck[2,3], R. J. C. Spreeuw[2,3], and C. Morais Smith[1#]

[1]*Institute for Theoretical Physics, Utrecht University, Princetonplein 5, 3584CC Utrecht, The Netherlands*
[2]*Van der Waals-Zeeman Institute, Institute of Physics, University of Amsterdam, Science Park 904, 1098XH Amsterdam, The Netherlands*
[3]*QuSoft, Science Park 123, 1098XG Amsterdam, The Netherlands*
[4]*Eindhoven University of Technology, P.O. Box 513, 5600MB Eindhoven, The Netherlands*
[5]*Departamento de Física Teórica e Experimental, Universidade Federal do Rio Grande do Norte, Campus Universitário, Lagoa Nova, Natal-RN 59078-970, Brazil*
[6]*Dipartimento di Fisica e Astronomia, Universitá di Firenze, I-50019, Sesto Fiorentino (FI), Italy*

Date: August 14 2025

\* These authors contributed equally to this work
\# Corresponding author. Email: c.demoraissmith@uu.nl

**Rydberg atoms trapped by optical tweezers have emerged as a versatile platform to emulate lattices with different geometries, in which long-range interacting spins lead to fascinating phenomena, ranging from spin liquids to topological states of matter. Here, we show that when the lattice has a fractal geometry with Hausdorff dimension 1.58, additional surprises appear. The system is described by a transverse-field Ising model with long-range van der Waals interactions in a Sierpiński gasket fractal. We investigate the problem theoretically using exact diagonalization, variational mean field, quantum Monte Carlo, and a graph-based numerical technique, SIM-GRAPH, which we developed. We find that in the quantum regime, the phase diagram exhibits phases in which the spins flip one-by-one. The theoretical results are in excellent agreement with experiments performed with single $^{88}$Sr atoms trapped by optical tweezers arranged in a fractal geometry. The magnetization and von Neumann entanglement entropy reveal several regimes in which single spin-flips are delocalized over many sites of one sublattice, thus allowing for an unprecedented control of a cascade of phase transitions in a many-body system. These results expand the possibilities of Rydberg atoms for quantum information processing and may have profound implications in quantum technology.**

Interactions play a prominent role in determining the ground state of quantum systems, but their influence is strongly dependent on the dimensionality of the system. Although in three-dimensions (3D) the effect of interactions can be cast into a renormalized band mass for quasi-free electrons, in 2D they may give rise to fractionally charged excitations in the presence of a perpendicular magnetic field [1], and in 1D they lead to Luttinger-liquid behavior with spin-charge separation [2]. Moreover, they may generate Majorana zero modes in 1D topological superconducting systems [3].

Most of the studies involving interactions are restricted to the Hubbard model, which accounts for local interactions [4-6]. Recent advances in cold atoms using Rydberg states, however, have allowed the realization of highly correlated systems with long-range van der Waals interactions [7-13]. Elusive states of matter have been experimentally emulated, such as quantum spin liquids [14], topological bosonic states [15], and strings in lattice gauge theories [16]. In particular, the transverse-field Ising model has been investigated in several geometries, ranging from 1D chains [17,18] to 2D arrays with square, honeycomb, and triangular symmetry [17-20], to 3D lattices [21]. Nevertheless, the role of *long-range interactions at non-integer dimensions,* as realized in fractals, has not yet been unveiled [22].

Fractals are structures that usually exhibit self-similarity and may have non-integer dimensions [23-28]. They can be found in the shape of rivers, coastal lines, as well as in the human body. Indeed, the circulatory system, the lungs, the intestines, and the neuronal system, are all fractals. Recently, electronic *quantum* fractals were artificially designed in the nanodomain [29]. Moreover, they were shown to emerge *spontaneously* in topological systems [30] and in the shape of proteins [31]. A paradigmatic example of a fractal, the Sierpiński gasket, which has a Hausdorff dimension d = 1.58, is shown in Fig.1a. Theoretical studies of quantum criticality in fractals have so far focused on the frustrated Sierpiński triangle (Extended Data Fig.1b) or carpet (Extended Data Fig.1e) including only nearest-neighbor (NN) interactions [32-35]. One very interesting feature of the Sierpiński lattice considered here is that it has sites with different connectivity, characterized as different sublattices (see colors in Fig.1b). Moreover, there may be an imbalance in the number of sites in each sublattice. These two effects together may lead to greater complexity and very interesting quantum states in the presence of interactions. Indeed, it has been recently shown that the local (Hubbard) interaction in a fractal leads to a long-sought metallic ferrimagnetic regime at half-filling [36], which never occurs in regular 2D lattices. In 2D geometries, the states at half-filling are either magnetic and insulating, or paramagnetic and metallic; magnetism never co-exists with metallicity.

Here, we investigate theoretically and experimentally the role of *long-range interactions* in a *fractal* geometry. We use optical tweezers to arrange single $^{88}$Sr atoms in a Sierpiński gasket fractal, see Fig.1c for a sketch of the experiment. Upon excitation with a laser that drives the transition between an atomic metastable (effective 'ground') state and a Rydberg excited state, the system then emulates an Ising model in a transverse field, described by the Hamiltonian

$$H = \sum_{0<i<j\leq N} V_{ij}\, n_i^z n_j^z + \sum_{i=1}^{N} \left(\frac{\hbar\Omega}{2}\sigma_i^x - \hbar\Delta n_i^z\right),$$

where $i$ and $j$ denote lattice sites, $V_{ij} = C_6/r^6 = V(a/r)^6$ is the long-range van der Waals interactions, $a$ is the lattice parameter, $n_i^z = (1 + \sigma_i^z)/2$, and $\sigma_i^x$ and $\sigma_i^z$ are Pauli matrices. The eigenstates of $n_i^z$ are the simulation ground state $|\downarrow\rangle$ and the Rydberg excited state $|\uparrow\rangle$, which mimic a spin−1/2 in a transverse-field Ising model, see sketch in Fig.1c. The detuning Δ acts as a longitudinal field and the transverse field arises from an ultraviolet (UV) 'Rydberg laser', driving the transition $|\downarrow\rangle \leftrightarrow |\uparrow\rangle$ with Rabi frequency Ω. The strong interactions can be expressed in terms of the Rydberg blockade radius $R_b$, the distance within which the excitation of another Rydberg atom is suppressed. It is related to $V$ as $V/\Omega = (R_b/a)^6$. The blockade mechanism competes with the transverse field Ω that flips the spins. Initially, all the atoms are in the spin-down state, see the bottom-left image in Fig.1c. The ground state of $H$ is prepared by means of adiabatic sweeps of the parameters (Ω, Δ) to their desired final value, see Methods for details. The final state is measured by taking a picture showing $|\downarrow\rangle$ atoms as present, $|\uparrow\rangle$ atoms as absent. Thus, when an atom is excited, it is no longer visible in the experiment, see bottom-right image.

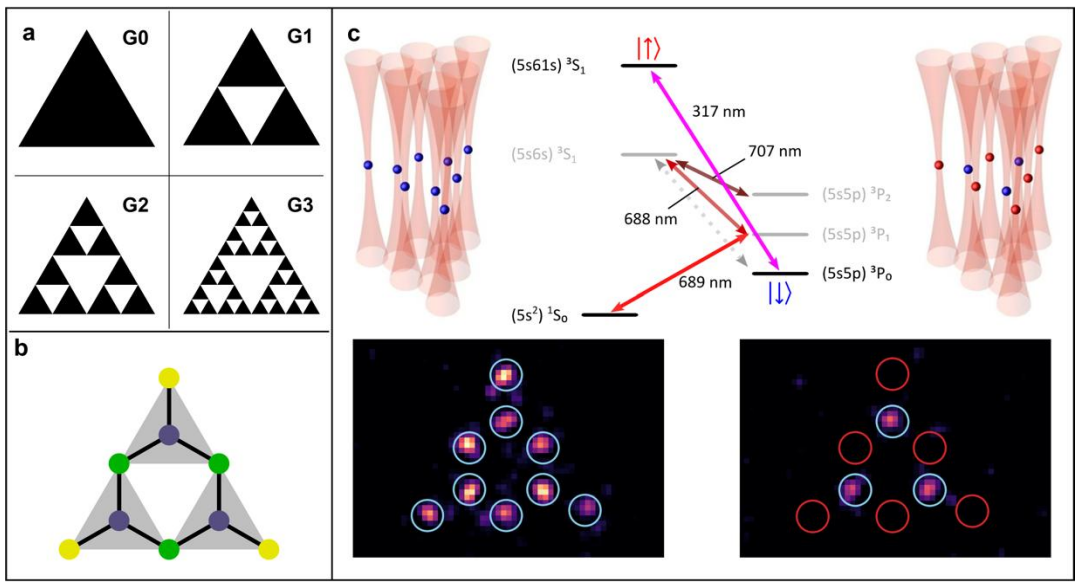

**Fig. 1| Sketch of a Sierpiński gasket and of the experimental setup**. **a**, Zeroth, first, second, and third generation of a Sierpiński gasket. **b**, Representation of one of the possible lattices for the first generation, in which sites are assigned to the vertices and to the centre of the filled triangles; see Methods for details on how to assign lattices to fractals. The sites are colored according to their connectivity: yellow corner sites have only one neighbor, green edge sites have two neighbors, and purple bulk sites have three neighbors. **c**, Top: A schematic depiction of the optical tweezer fractal arrangement used in the experiments, together with the relevant level scheme for [88]Sr. The simulation ground (down, blue) and excited (up, red) states are coupled with 317-nm light. Bottom: Fluorescence images of atoms before and after a simulation. The atoms in (down) are detected while the atoms in (up) are absent.

We start by investigating the ground-state magnetization $\langle m \rangle = (\langle n_\uparrow \rangle - \langle n_\downarrow \rangle)/N$ using exact diagonalization (ED) for the first-generation Sierpiński gasket, which is composed of 9 sites. Here, $\langle n_\uparrow \rangle = \sum_{i=1}^{N} \langle n_i^z \rangle$, where $N$ is the total number of sites, $\langle n_\downarrow \rangle = N - \langle n_\uparrow \rangle$, and $P_i[\uparrow] = \langle n_i^z \rangle$, the spin-up probability at site $i$. The transverse-field Ising model is studied in this fractal lattice for different values of the interaction V, detuning Δ, and transverse field Ω.

**Magnetization: classical case**
In the absence of a transverse field, the system is in a classical regime, and correlations are not important. In this case, a different number of phases appear in the ground-state phase diagram, depending on whether only NN or long-range interactions are considered. For NN interactions, there is a phase with all spins down, an antiferromagnetic phase with three spins down and six up, and a ferromagnetic phase with all spins up, see Fig.2a. Phases with intermediate numbers of spin flips may occur in the presence of *long-range interactions*. Consider the phase with three spins up in Fig.2b. Initially, the spins up appear at the corners of the triangle, to minimize the long-range interaction. Then, if the parameters are tuned such that it is energetically favorable to have four spins up, the additional spin will sit on a single edge site (green sites in Fig.1b). These classical states are three-fold degenerate, and at each realization, the fourth spin up will be detected in a random green edge site. Upon increasing the detuning $\Delta$ further, there will be two spins up in two of the three edge sites, in addition to the three spins up at the corners. For even larger values of $\Delta$, a stable antiferromagnetic phase with six spins up becomes the ground state. Finally, for very large detuning, the spins flip again one by one, but now in the bulk (purple sites in Fig.1b), until all the nine spins are up.

**Magnetization: quantum case**
On the other hand, when the transverse field $\Omega$ is finite and quantum effects become relevant, the phase diagram becomes much richer. In the presence of only NN interactions, the phases are the same as in the classical case (Fig.2c), but with van der Waals interactions, regimes with 3, 4, and 5 spin-flips become visible for a larger regime of parameters, and there is a sharp boundary separating phases with a different number of spins up, see Fig.2d. The spin configuration in each phase at the parameter values corresponding to the black dots in Fig.2d are displayed in Fig.2e. One very interesting feature emerges: for the phase with four spin flips, there are three spins up at the corners and the additional spin up delocalizes in a *superposition state*, on all equivalent edge sites (Fig.2e). For larger values of $\Delta/\Omega$, there are five spins up, and two spins are in a superposition state on the edge sites. This is in contrast with the classical case, where the additional spin up for $\langle n_\uparrow \rangle = 4$, e.g., sits on a single site. Upon increasing the value of $\Delta/\Omega$ even further, the antiferromagnetic state with six spins up is reached. These results for the long-range interacting quantum fractal indicate that it is possible to *control and access sublattice spin-flips one by one*, since these superposition states appear in a broad regime of parameters.

**Comparison between theory and experiments**
Our theoretical findings obtained using ED for the first-generation Sierpiński gasket are corroborated by experimental measurements of the same phases, see Fig.2g. One observes a very good agreement between theory and the raw experimental data. Upon including fluctuations in the position of the atoms, which more realistically describe the experimental conditions, the theoretical results qualitatively agree even better with the data, see Figs.2f,g. Small fluctuations may slightly break the degeneracy between sites of the same kind and lead to a stronger localization in one or two of them, but the additional spin up remains in a superposition and the quantum state is not destroyed. In Methods and Extended Data Fig.4, we present the same data using an extreme color code to emphasize small variations in the excitation probabilities. In addition, we theoretically investigate the role of defects in the position of the atoms and include a SPAM correction of the raw experimental data, which leads to an even better agreement with theory.

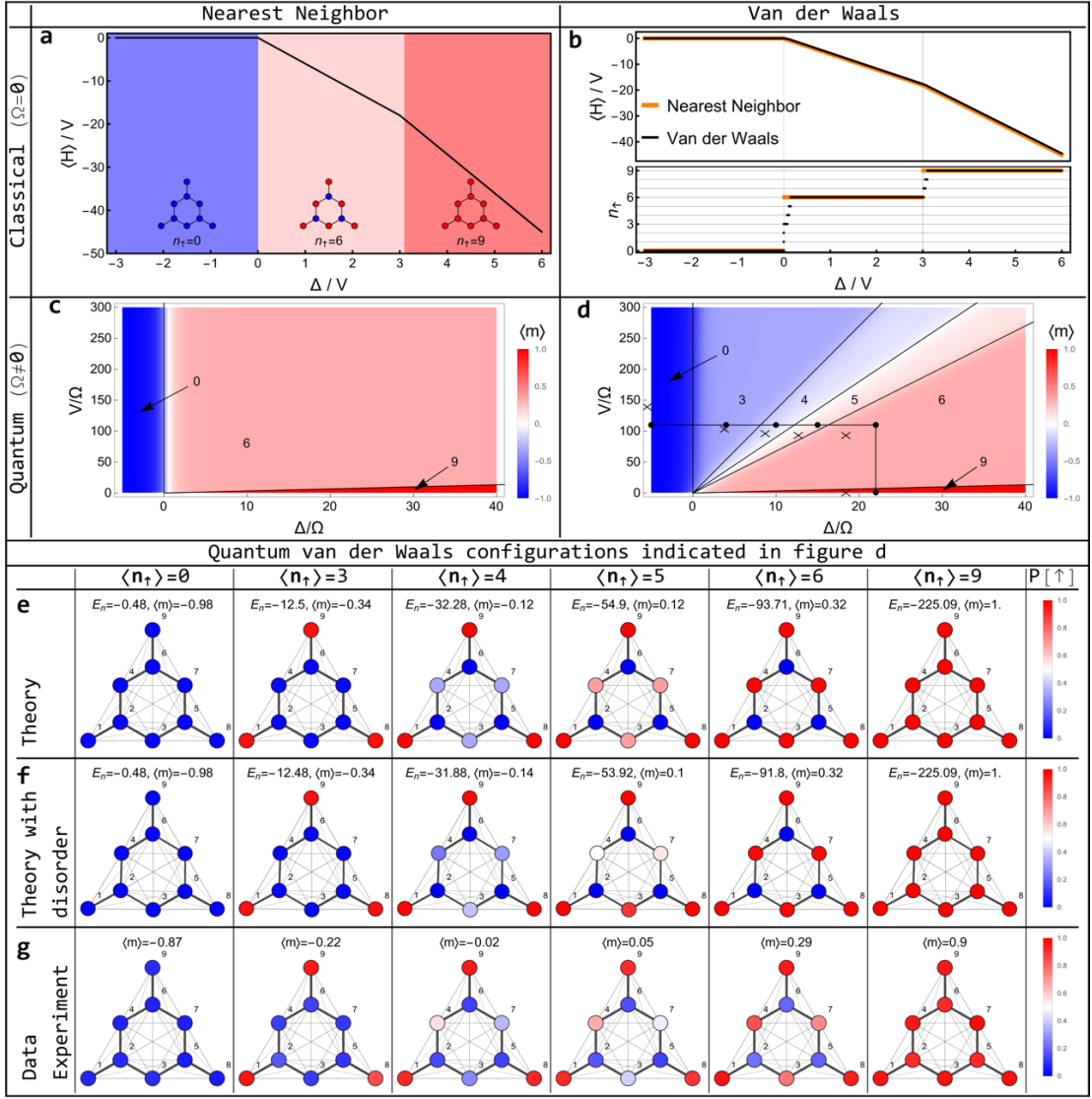

**Fig. 2| Phase diagrams obtained using ED in the classical ($\Omega = 0$) and quantum ($\Omega$ finite) regimes, for NN and van der Waals interaction; spin configurations obtained theoretically and experimentally. a,b** Magnetic phase diagrams for the classical case. When considering only NN interactions **(a)**, states with zero, six and nine excitations are visible. Additional states emerge with van der Waals interaction **(b)**, but in a very small regime of parameters. **c,d** Magnetic phase diagrams for the quantum case with NN **(c)** and van der Waals **(d)** interactions. For high interactions, superposition states with four and five excitations become apparent in **(d)** in a broad regime of parameters. **e,f,** Spin configurations calculated for the parameters indicated by black dots in **d,** corresponding to $\Delta/\Omega$ = -4.5, 4.1, 10, 15, 21.8 with $V = 110\,\Omega$, and $\Delta/\Omega$ = 21.8 with $V = 0$, respectively. The calculations in **f** include random fluctuations in the position of the atoms, taken from a Gaussian distribution with a standard deviation of 1% of the NN distance. **g,** Experimental observations at the crosses in **d**. The first cross is out of the region shown in the phase diagram. Calculations performed at the crosses yield the same results as at the dots, thus confirming the robustness of the quantum phases.

## Higher-generation fractals

We now explore higher generations of the Sierpiński gasket numerically. As the system size increases, the complexity doubles in size (~$2^N$) for every new particle added, and conventional numerical techniques become very costly [37]. Therefore, we have developed a method based on graphs and symmetries, named SIM-GRAPH, standing for Symmetric Ising Models-Graph Reduction And Projected Hamiltonians. It avoids much of this complexity by projecting symmetric sites into a single site, effectively reducing $N$ by the symmetry factor of the geometry. In Fig.3a, we present the magnetization obtained for the second generation Sierpiński gasket using SIM-GRAPH (colours) and variational mean field (VMF), dashed lines. Up to the symmetry factor of SIM-GRAPH, we find good agreement between them (see Methods for limitations and advantages of SIM-GRAPH). In Figs.3b,c, we compare the results obtained with VMF, SIM-GRAPH, ED, and quantum Monte Carlo (QMC) along the black-solid line in Fig.3a, and in Figs.3d,e we present the spin configurations. Figure 3 confirms our findings for the first generation, namely that in the presence of long-range interactions, one can access regimes in which the number of spin flips changes one-by-one. Indeed, for the second generation, we find plateaus for $6 \leq \langle n_\uparrow \rangle \leq 15$. The state with $\langle n_\uparrow \rangle = 7$, e.g. contains four spins in a superposition on twelve sites along the inner edge of the gasket. The state with $\langle n_\uparrow \rangle = 12$ is an exception: SIM-GRAPH predicts a very stable plateau, but the other techniques show that this state is metastable [38]. Nevertheless, the computational time required for the four methods is very different: from several days (hours) per point of ED (QMC), several server hours for Fig.3b with mean-field, and less than 30 seconds on a laptop for SIM-GRAPH. Keeping this in mind, we emphasize the exceptional accuracy of SIM-GRAPH near low $V/\Omega$ and low $\Delta/\Omega$ (insets of Figs.3b,c), where the method is expected to function well.

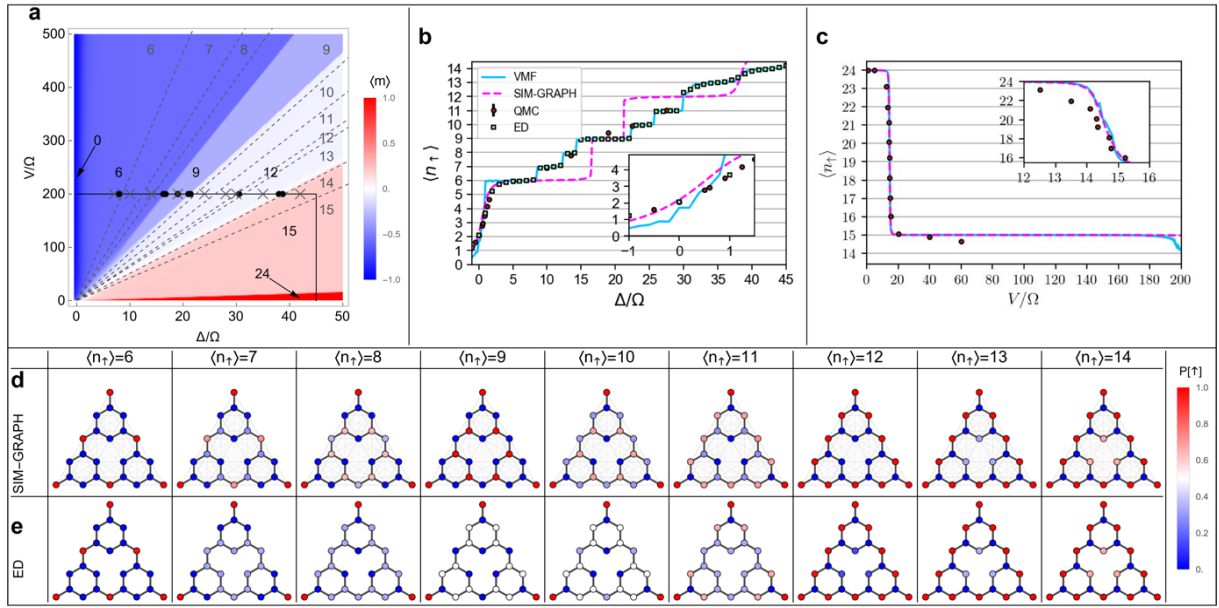

**Fig. 3| Second-generation Sierpinski gasket magnetization and spin configurations. a,** Magnetization phase diagram calculated with SIM-GRAPH (colors, $\langle n_\uparrow \rangle$ in black) and VMF (dashed lines, $\langle n_\uparrow \rangle$ in grey). **b, c,** Comparison of results across the horizontal and vertical black line in **a**, respectively. **d, e,** The spin configurations predicted by SIM-GRAPH and ED, computed at the black dots and grey crosses in **a**, respectively. The antiferromagnetic ($\langle n_\uparrow \rangle = 15$) and ferromagnetic ($\langle n_\uparrow \rangle = 24$) phases are not shown.

Results for the third generation obtained using SIM-GRAPH are presented in Extended Data Fig. 5. For this system size, the other numerical techniques are excessively costly. The phases change in units of 3, the symmetry factor of the gasket, but inspection near the phase transitions can reveal spin configurations in which the spin flips one by one, as observed before for the first and second generations.

**Correlations** Next, we turn our attention to correlation effects and calculate the spin susceptibility and the entanglement entropy of the first-generation Sierpiński gasket. One can obtain the phase diagram by identifying peaks of the spin susceptibility $\chi = -\partial^2 E_0 / \partial \Delta^2$ [39], which determines the variation of the ground state energy $E_0$ as a function of the longitudinal field $\Delta$. In Figs.4a,b, we present the results obtained upon tuning the long-range interaction, displayed in terms of $V$ and the Rydberg blockade radius $R_b$, respectively. Alternatively, we can calculate the bipartite von Neumann entanglement entropy $S_{vN}$ of the ground state, which is defined by considering the reduced density matrix $\rho_r$ of one of the sub-systems (sites 1-4 in Fig.2e), $S_{vN} = -\text{Tr}[\rho_r \ln(\rho_r)]$, see Figs.4c,d. The phase diagrams shown in Figs.4a,c corroborate the magnetization results presented in Fig.2d. The ordered phases (black regions in the phase diagram), which result from the interplay of the blockade mechanism with the detuning, have small entropy. The high-entropy disordered phases (red region in Fig.4d) may host the elusive spin-glass or spin-liquid states.

Now, we further compare theory and experiments. One way to identify the phases is to use the correlations to calculate the structure factor (see Methods) and the momentum distribution in the reciprocal space. However, momentum is not a good quantum number because fractals lack translational symmetry. Nevertheless, a Fourier transform can always be performed, and it was shown that the self-similarity is inherited in $k$-space [29]. In Figs.4e,f, we present the theoretically calculated and the experimentally measured spin configurations, respectively. We focus on the states with zero, three, four, five, and six spin flips, and compare theory (Fig.4e,g) and experiments (Fig.4f,h) by analyzing the correlation functions and their Fourier transform. Since the states with one and two spin flips are not number states, we do not show them. The reciprocal space allows one to understand better how the different length scales corresponding to the different sublattices emerge upon tuning the parameters of the problem. Initially, the system is in the zero spin up phase, and there are well separated peaks in Fourier space, given by the inverse of the smallest length scale in the problem, the lattice spacing (first panel in Figs.4g,h). This behavior changes as three spins are flipped at the corners of the gasket, when the size of the Sierpiński triangle emerges as a new length scale, corresponding to a very dense lattice in $k$-space. The third length scale is an intermediate one, given by the size of the missing triangle (green sublattice in Fig.1b), which induces an intermediate-length structure in $k$-space (see third panel in Figs.4g,h). This phase initially coexists with the dense phase when the fourth spin is flipped, but it becomes more dominant as the system approaches the antiferromagnetic ground state. When more than six spins are flipped, the bulk sites (purple sublattice in Fig.1b) start to play a role, but the length scale of this sublattice is the same as the green one, and there is no fundamental modification of the structure in $k$-space, except that now this structure will evolve again towards the very large one in $k$-space, corresponding to the inverse of the lattice constant, which is probed when all spins have flipped up. Therefore, we restricted the analysis to the cases of spin-flip up to six. There is an excellent agreement between theory and the raw experimental data.

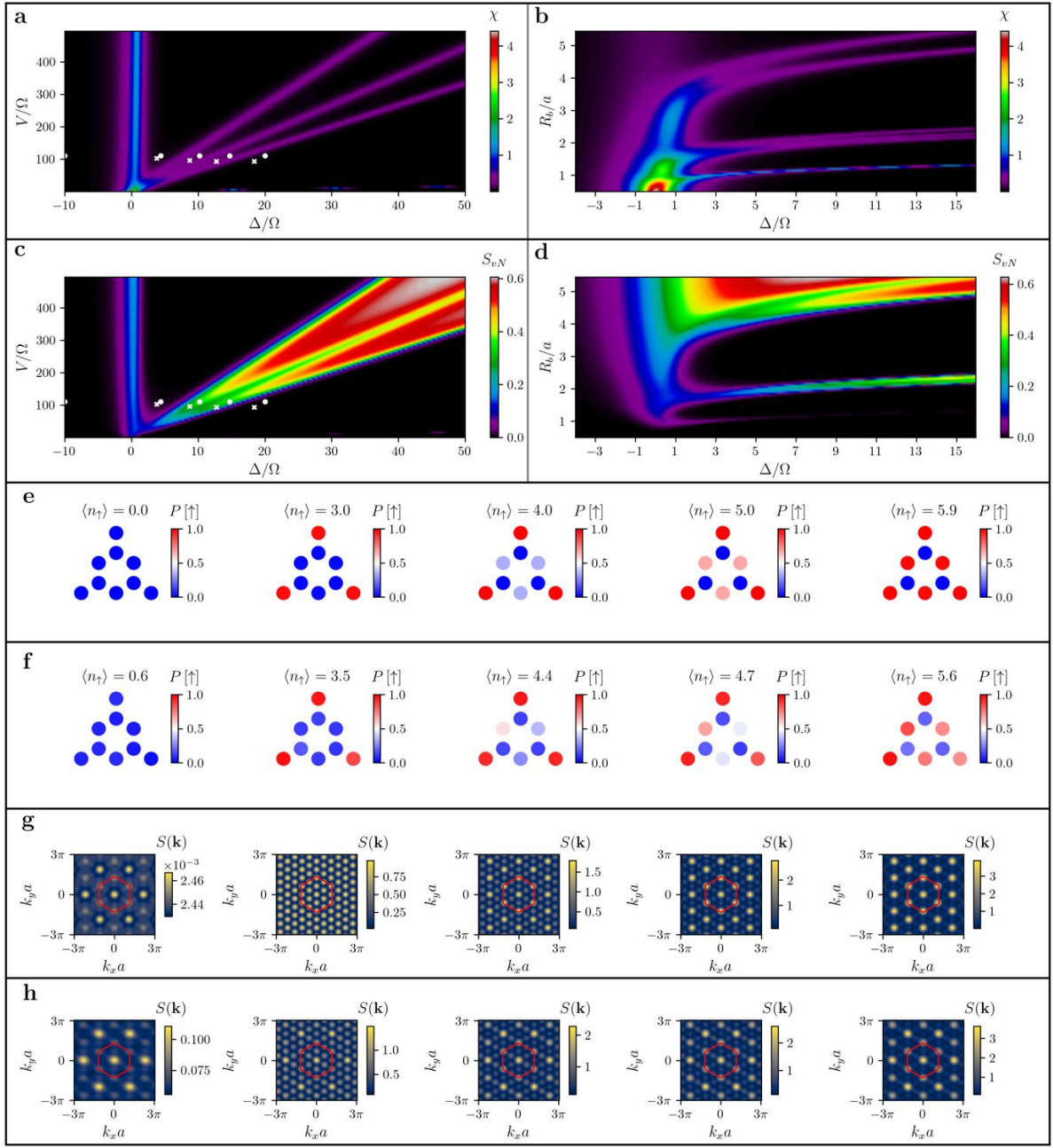

**Fig. 4| Spin Susceptibility and von Neumann entanglement entropy. a,b,** Spin susceptibility $\chi$ and **c, d,** von Neuman entanglement entropy $S_{vN}$ as a function of $V/\Omega$ and $R_b/a$, respectively, versus $\Delta/\Omega$. **e,** Spin configurations obtained from ED and **f,** measured experimentally in the regime of spin flips from 0-6; the states with 1 and 2 excitations are not shown because at this interaction strength, these are not number states. **g,** Correlation functions calculated numerically along the white dots and **h,** obtained experimentally at the white crosses shown in **(c)**. Calculations performed at the crosses yield the same results, which emphasizes the robustness of the states. The points in $k$-space are extended beyond the red hexagon to help visualization.

The interplay between the long-range Rydberg interaction and the different sublattices in the fractal leads to an unprecedented control of the sublattice scales in parameter space, and a clear signal in the structure factor. The tuning of the parameters $V$, $\Delta$, and $\Omega$, which are experimentally accessible, defines which sublattice is probed and which one is more relevant in each regime of parameters. This is fundamentally different than spin flips in regular lattices, in which all sites are equivalent. Our results thus reveal that fractality, together with long-range interactions, open unexpected directions of control and manipulation of quantum phases in Rydberg atom structures. Regimes in which there is a single-spin flip in a quantum superposition on sites of the same sublattice are shown to be experimentally accessible. Moreover, the order in which the different sublattices emerge is determined by the rich interplay of the range of the interactions and the geometry of the fractal.

As an outlook, it would be interesting to understand whether there is any connection between the Ising model in a transverse field and Majorana zero-modes at fractal dimensions, as there is in 1D [40]. In addition, the realization of Heisenberg spins in a dual Sierpinski gasket as depicted in Extended Data Fig.1b, which is triangular, could reveal how fractality could affect the spin-liquid phase, especially in the presence of long-range interactions. Moreover, the combination of fractality, topology and long-range interactions remains to be investigated. It was recently shown that fractality per se can induce topological states [41]. It would be important to understand how long-range interactions affect these phases. Furthermore, a 2D XY-spin model on a square lattice was recently realized for Rydberg atoms with dipole-dipole interactions, and a phase transition due to the breaking of a continuous symmetry was revealed [42]. An experimental setup in which a fractal with XY spins could be combined with a quantum simulation of spin-orbit coupling and long-range interactions could allow the realization of a compelling case. Finally, it would be paramount if one could use the controllable single-spin-flip phenomenon discussed here for quantum sensing and quantum information processing.

**METHODS**

**1. Theory**

**Fractals.** Between the periodic structure of a lattice and the chaos of randomness, quasi-crystals and fractals show an intermediate level of structure. Whereas quasi-crystals can be mapped to higher-dimensional lattices, fractals are structured according to their scale invariance. That is, upon scaling the structure, one retrieves either the exact same (for a self-similar fractal) or a structure akin to the original in a statistical sense. Self-similar fractals can often be constructed using a repeating scheme, adding or subtracting specific sections again and again, where each step is called a generation [26]. Usually, this results in a very fine detail within the structure, with many kinks or holes, up to a finite size. The scaling behaviour of fractals can be captured using the Hausdorff dimension [24], which can be non-integer.

An important question arises when we consider quantum particles on a fractal, as we need to define a lattice. Some of the options for the Sierpiński gasket and carpet are shown in Extended Data Fig. 1, where one selects either the corners, centres, or both from the unit cells of the fractal (shown in blue). In the literature, the centre lattice is also called the dual lattice, whereas the corner-lattice is typically called the triangular (resp. square) lattice, and the combined lattice is called the hexagonal (resp. diamond) lattice. These choices can result in vastly different lattices, but there are some common features. The corner-only choice consistently misses the smallest holes of the fractal, effectively losing one or even two generations. The centre points make a very minimal cover of the fractal, but the connections show that the sizes of the holes (related to lacunarity) can vary significantly compared to the original fractal, altering the geometry. Finally, the combination of both presents a reasonable balance between enough bulk and connectivity which preserves the holes. Although this comes at the cost of requiring the largest number of lattice points, it is still the most efficient lattice since it never loses a generation. This is the lattice that we consider in this work.

**Ising model on a Graph.** A graph consists of a set of vertices (points) that are connected by a set of edges (lines) [43]. In general, a graph is independent of the location of the vertices and is purely defined based on its connectivity. When considering an Ising Hamiltonian, all the terms can be captured by introducing a graph on the geometry. After labelling all positions of the spins as a vertex, we can model the external fields $\Delta$ and $\Omega$ as self-looping edges $i-i$ with the correct edge-weights, while interactions can be modelled by edges between the different vertices $i-j$ that are weighted by the interaction strength $V_{ij}$. The weighted connectivity matrix of the graph can be used to reconstruct the corresponding Hamiltonian. By considering a binary basis of states in the z-axis, we can cast the Hamiltonian into a matrix of size $2^N \times 2^N$. Here, the detuning and interaction terms fall onto the diagonal, and the off-diagonal contains the transverse field, forming a repeating patten of off-diagonal identity matrix blocks. To better understand the high-dimensional wave-vectors, we have plotted them with a heat-map according to the local spin-up probability $P_i[\uparrow] = \langle n_i^z \rangle$.

**Exact Diagonalization.** To perform ED, we first choose a suitable basis of states $\{|n_1 n_2 \ldots n_N\rangle\}$, with $n_i = 0$ or 1, to write the Rydberg Hamiltonian in its matrix format. This basis has $2^N$ elements, resulting in a $2^N \times 2^N$ matrix to represent the Hamiltonian. The exponential scaling with the system size makes the numerical diagonalization of the Hamiltonian only tractable for small systems, such as the first generation of the Sierpiński triangle. For the second generation, we only perform the calculations at selected points of the phase diagram.

**Variational Mean-Field.** In the mean-field description, we exclude quantum correlations and approximate the ground state of the system using a variational product-state wave function

$$|\Psi_V\rangle = \bigotimes_{i=1}^{N} (\cos\theta_i |0\rangle + \sin\theta_i |1\rangle),$$

where $\theta_i$ are parameters. The variational principle states that the energy associated to $|\Psi_V\rangle$,

$$E_V = \frac{\langle\Psi_V|H|\Psi_V\rangle}{\langle\Psi_V|\Psi_V\rangle} = \sum_{i<j} V_{ij} \sin^2\theta_i \sin^2\theta_j$$
$$+ \sum_i \left(\frac{\Omega}{2}\sin 2\theta_i - \Delta \sin^2\theta_i\right),$$

is an upper bound to the ground-state energy $E_0$. In that sense, the natural procedure is to minimize $E_V$ with respect to the variational parameters $\theta_i$ to make it as close as possible to $E_0$. Since the number of variational parameters increases linearly with the system size, we perform this minimization numerically using the limited-memory Broyden-Fletcher-Goldfarb-Shanno algorithm [44].

**Quantum Monte Carlo.** The QMC technique employed in this work is the Stochastic Series Expansion method for spin systems [45], which has recently been adapted to simulate the Rydberg Hamiltonian [46]. The core of the method is the Taylor expansion of the partition function,

$$Z = Tr[e^{-\beta H}] = Tr\left[\sum_{n=0}^{\infty} \frac{\beta^n}{n!}(-H)^n\right],$$

where $\beta$ is the inverse temperature, often referred to as imaginary time. To perform the trace, it is natural to choose the Rydberg occupation basis $\{|\alpha\rangle\} = \{\bigotimes_{i=1}^{N} |n_i\rangle\}$, with $n_i = 0$ or $1$. Therefore, the partition function becomes

$$Z = \sum_{\{\alpha_p\}} \sum_{n=0}^{\infty} \frac{\beta^n}{n!} \prod_{p=1}^{n} \langle\alpha_{p-1}|-H|\alpha_p\rangle,$$

with $|\alpha_0\rangle = |\alpha_n\rangle$. The next step is to break the Hamiltonian in a sum of different terms, $-H = \sum_{t,a} H_{t,a}$, where $t$ indicates the type of operator (diagonal or off-diagonal) and $a$ indicates the place where the operator acts on the system (sites or links between sites). After this step, the partition function becomes

$$Z = \sum_{\{\alpha_p\}} \sum_{n=0}^{\infty} \frac{\beta^n}{n!} \prod_{p=1}^{n} \sum_{t,a} \langle\alpha_{p-1}|H_{t_p,a_p}|\alpha_p\rangle$$
$$= \sum_{\{\alpha_p\}} \sum_{n=0}^{\infty} \sum_{S_n} \frac{\beta^n}{n!} \prod_{p=1}^{n} \langle\alpha_{p-1}|H_{t_p,a_p}|\alpha_p\rangle,$$

where $S_n$ is a particular sequence of the operators $H_{t_p,a_p}$ in the product. With this expression for the partition function and a truncation of the summation in $n$ for $n > n_{max}$, we have the ingredients to perform a QMC simulation. The probability density function (PDF) to be sampled is given by the matrix elements $\langle\alpha_{p-1}|H_{t_p,a_p}|\alpha_p\rangle$, which must be positive to avoid a sign problem. For the Rydberg system, some transformations are applied to the Hamiltonian to fulfil this condition, see details of these transformations and how to sample the PDF in Ref. [46]. After these transformations, we define the types

$t$ and locations $a$ of each Hamiltonian element for our system. Specifically, we have the diagonal operators $t = 1$, off-diagonal operators $t = -1$, operators acting on lattice sites $a = s$, operators acting on the links between the sites $a = l$, and the identity operator $H_{0,0} = 1$. The other terms are: $H_{-1,s} = (\Omega/2)\sigma_x$; $H_{1,s} = (\Omega/2)\mathbf{1}$; and $H_{1,l} = -V_{ij}n_i^z n_j^z + \delta(n_i + n_i^z) + C_{ij}$, where $\delta = \Delta/(N-1)$, $C_{ij} = |\min(0, \delta, 2\delta - V_{ij})| + \epsilon|\min(\delta, 2\delta - V_{ij})|$, and $\epsilon > 0$ is a simulation parameter.

The calculation of thermodynamic properties is done using derivatives of the partition function; for instance, the energy is $E = -\partial \log Z / \partial \beta$. The Rydberg density is calculated by counting the number of excited atoms in each sample. To obtain ground-state estimates, an extrapolation to $\beta \to \infty$ is needed. We performed simulations using the SSE method implemented in the Bloqade software [47].

**SIM-GRAPH.** We developed a method named SIM-GRAPH, which uses symmetry and graph reduction to construct a projected Hamiltonian on a reduced system. The method is designed to use symmetries in Ising models in finite lattices lacking translational symmetry, to reduce the size of the Hamiltonian matrix. This is done by first defining the unique symmetric points in the graph representing the interactions, and then projecting all interactions onto this subset of points. Effectively, we reduce a simple large graph to a complex small graph. This reduced graph defines a new Hamiltonian, which models all symmetric states of the original system. Depending on geometry, this procedure might miss some anti-symmetric states, but the observables will always be symmetric. Thus, the goal is not to reproduce the full wavefunction, but to reproduce the observables. The strength of this method is on the reduction of the number of spins before constructing a Hamiltonian matrix. It avoids the exponential $2^N$ scaling for large $N$, and instead uses a reduced size $N_S$, the value of which depends on the geometry.

We begin by calculating the automorphism group of the graph of interactions, whose cycles reveal the different equivalent points. Selecting, for example, the most bottom-left point from each class yields a set of symmetry points $S$ (Extended Data Figs.2a and c). We can now map all edges of the original interaction graph onto the symmetric points, by projecting them to the symmetry point within each of the classes, resulting in the reduced interaction graph (Extended Data Fig.2b). Using the same techniques as before, the reduced graph is converted into a matrix, which is no longer of size $2^N$, but of size $2^{N_S}$, where $N_S$ is the number of points in $S$. Due to this exponential reduction in matrix size, ED remains usable for much larger system sizes.

There are some restrictions to this method, which come from our assumption that similar states will occupy symmetric sites equally. Firstly, this is only true up to the point where the Rydberg blockade radius forbids two sites to be both up. Thus, once $R_b$ reaches the size of the symmetry reduced graph, $L_S$, we will find less spin-up electrons on this sub-graph, whereas there should more spin-up electrons on the full graph. Hence, in principle, SIM-GRAPH is limited to $R_b \lesssim L_S$. Secondly, SIM-GRAPH works best on odd geometries, where the symmetry axes cut through the sites instead of between the sites, since this avoids the possibility of missed superposition states on the boundary of the reduced graph.

To demonstrate the effectiveness and limitations of SIM-GRAPH, we compare the phase diagram of the first-generation Sierpiński gasket calculated using ED and SIM-GRAPH (Extended Data Fig.3). We use an extreme colour code to emphasize the differences in the spin configurations. Nevertheless, we find that the results agree almost everywhere, except for the two superposition states, which have collapsed onto either side of the transition from 4 to 5. The SIM-GRAPH results only show stable regions for the symmetric configurations of 0, 3, 6 and 9. Thus, we conclude that SIM-GRAPH effectively predicts the magnetization up to the symmetry factor of the geometry but is able to capture the missing spin configurations (4 and 5 spins up) at the boundaries of the phase transitions.

**Disorder and defects.** In Extended Data Figs.4a and b, we repeat the theoretical and raw experimental results shown in Fig. 2 of the main text, but now using an extreme colour-scale to highlight the differences. In addition, in Fig.4c we show the SPAM corrected experimental data (see Experimental setup below for details). Some of the experimental results break the theoretical symmetry of the model. They can be explained by a small amount of disorder and defects. The disorder is described by a positional Gaussian noise with a standard deviation σ = 1% of the NN distance (Extended Data Fig.4d). To model a defect, the position of site 4 is moved outwards by 10% of the NN distance (Extended Data Fig.4e). We find that only the two superposition states (4 and 5) seem to be significantly affected by disorder and defects. Although the superposition state is no longer symmetric in the presence of disorder or defects, it remains robust, with an asymmetric amplitude on the different sites of the same sublattice.

## 2. Experimental setup

### A. Experimental Sequence

**Defect-free single-atom array.** All experiments start by stochastically (p~50%) loading a pattern of 6×6 optical tweezers ($\lambda$=813 nm) with single $^{88}$Sr atoms and detecting the atoms by collecting fluorescence during illumination with 689-nm light for 200 ms, as described in detail in Refs. [48-51]. All experiments are performed at magnetic fields zeroed to $B$<10 mG. The optical tweezers are generated with a phase-only spatial light modulator (SLM) with a >1 kHz refresh rate. The holograms on the SLM are updated every ~2.75 ms to simultaneously move nine selected tweezers and form the first-generation Sierpiński gasket. Before the move, unused and empty tweezers are extinguished, which redistributes their laser power over the nine selected tweezers. This results in deeper traps (~500 μK) and higher imaging survival and detection fidelities [50]. After rearrangement, a verification image is taken. In about 75% of the experimental runs, a defect-free 9-atom pattern is obtained.

**Optical pumping to the simulation ground state.** To populate the ground state ($|\downarrow\rangle$) of our simulation, the atoms are optically pumped to the 5s5p $^3P_0$ state by shining simultaneously 689-nm, 688-nm, and 707-nm light onto the atoms for 30 ms. In this way, atoms are pumped via the 5s$^2$ $^1S_0$ → 5s5p $^3P_1$→ 5s6s $^3S_1$ transitions, after which they decay to any of the 5s5p $^3P_J$ states, with J = 0,1,2. The 688-nm and 707-nm light excite decayed atoms from 5s5p $^3P_1$ and 5s5p $^3P_2$ back to 5s6s $^3S_1$, respectively, such that atoms accumulate in 5s5p $^3P_0$. This optical pumping leads to heating of the atoms because they scatter many photons (on average > 10) and experience different trapping potentials in different states; most notably, 5s6s $^3S_1$ is anti-trapped by 813-nm light. We characterize the heating by blinking the optical tweezers for a variable duration and measuring the recapture probability [49, 52]. We quantify the heating by comparing the result to classical Monte Carlo simulations for a trap depth of 500 μK and a tweezer waist of 0.88 μm to estimate an initial temperature of 10.8 μK.

**Quasi-adiabatic sweep.** After the atoms have been pumped to 5s5p $^3P_0$, we trigger an arbitrary waveform generator (AWG; Spectrum M4i.6622-x8) with 625 MSamples/s output rate, which produces two outputs on two channels: it produces a digital control voltage for the radio-frequency (RF) switch on the 813-nm beam path and it produces the RF signal used for ramping the 317-nm light for the Rydberg excitation. The control voltage extinguishes the tweezers during the whole UV excitation to avoid heating and atom losses by the anti-trapping of the 5s61s $^3S_1$ state at 813 nm [53]. To ensure the tweezers are fully extinguished before the UV light is present, there is a delay time of 1.2 μs between turning off the tweezers and starting the UV ramp. Likewise, after the UV ramp has finished, there is a 1.1 μs delay before the tweezers are turned back on. The total duration that the tweezers are extinguished is maximally 9.3 μs. Repeating the blink measurement from the previous paragraph for this duration, we measure an average survival of atoms in 5s5p $^3P_0$ of 95.8%, see SI for full error analysis.

To realize the Hamiltonian, Eq.1, atoms are illuminated by UV laser light controlled by the second channel of the AWG, see also UV laser system. Ramps are performed by changing the amplitude and frequency of an RF waveform sent to a 200-MHz acousto-optic modulator (AOM; Gooch & Housego 3200-1210) in a double-pass configuration (switching AOM). To minimize third-harmonic distortion, the programmed AWG amplitude is kept low ($V_{pp}$~100 mV), and two stages of amplification (Minicircuits ZFL-1000LN+ and ZHL-03-5WF+) are employed.

All RF ramps consist of five stages, see Extended Data Fig.6a. In the first part of the ramp, the frequency (blue trace) is kept constant at far red detuning from resonance, while the RF amplitude (red trace) is ramped on quadratically. Next, at constant amplitude, the frequency is swept from its initial to its final value in three steps: In the first part, $t_i \leq t_-$, the frequency increases following a third-order polynomial from $f_i$ to $f_-$. For $t_- \leq t_+$, the frequency grows linearly from $f_-$ to $f_+$. When $t_+ \leq t < t_f$, the frequency again grows polynomially from $f_+$ to $f_f$. At the end of the ramp, the frequency is again kept constant, and the amplitude is ramped down. In Extended Data Table1, the exact times and frequencies for each dataset used in the main text can be found.

**Final state Detection.** Upon turning back on the tweezers, the atoms in 5s5p $^3P_0$ are recaptured, while the atoms in 5s61s $^3S_1$ are expelled, leading to Rydberg atoms being detected as atom loss. Before imaging, we remove atoms that have decayed from 5s61s $^3S_1$ to 5s5p $^3P_2$ by repumping that state with 707-nm light and then blowing 5s$^2$ $^1S_0$ atoms away with a pulse of 461-nm light. There is a 25% chance during this process that atoms from 5s5p $^3P_2$ decay via 5s6s $^3S_1$ to 5s5p $^3P_0$, which we cannot distinguish from atoms that ended in 5s5p $^3P_0$ following the simulation. We estimate that this results in a false positive rate that is approximately 1%. The atoms in 5s5p $^3P_0$ are then pumped back to 5s$^2$ $^1S_0$ by illumination with 679-nm (see dotted grey lines in Fig.1c) and 707-nm light, and a 200-ms-long fluorescence image is taken. This image is compared to the verification image after rearrangement to obtain a raw detection probability of atoms surviving the Rydberg excitation. The raw detection probability contains contributions from all state-preparation and measurement (SPAM) errors.

Using the SPAM correction procedure described below, the raw probability can be used to calculate a corrected survival rate for the Rydberg excitation; however, only raw probabilities are reported in the main text.

**SPAM Correction Procedure.** To account for SPAM errors, we consider a simplified model of our experimental sequence. Each atom has a probability $p \approx 0.989$ to have been successfully pumped into 5s5p $^3P_0$ and we assume a false positive probability $\varepsilon_p \approx 0.048$ of detecting a Rydberg excitation when there was none and a converse false negative probability $\varepsilon_n \approx 0.01$. The values of these probabilities are obtained in calibration measurements that are presented in the SI. Combining these probabilities, we express a probability vector $\vec{P} = \begin{pmatrix} P_1 \\ P_0 \end{pmatrix}$ of detecting an atom as:

$$\vec{P} = p\boldsymbol{M}\vec{P}_\psi + (1-p)\vec{O}.$$

Here, $\boldsymbol{M} = \begin{pmatrix} 1-\varepsilon_p & \varepsilon_n \\ \varepsilon_p & 1-\varepsilon_n \end{pmatrix}$ is a matrix that determines how the projected probability $\vec{P}_\psi = \begin{pmatrix} \langle \downarrow |\psi\rangle \\ \langle \uparrow |\psi\rangle \end{pmatrix}$ of the atomic state following the Rydberg excitation is detected, and $\vec{O}$ is an offset vector that accounts for atoms that were not optically pumped. Assuming equal and independent errors for all atoms in a single simulation, this can be extended to $N$ atoms:

$$\vec{P}^N = p^N \boldsymbol{M}^N \vec{P}_\psi^N + (1-p^N)\vec{O}^N.$$

Now, $\vec{P}^N$ denotes the probability distribution of detected outcomes (*e.g.*, "11...11"), $\vec{P}_\psi^N$ the inferred probabilities of product states (*e.g.*, $|\uparrow\uparrow \cdots \uparrow\uparrow\rangle$) and $\boldsymbol{M}^N = \boldsymbol{M} \otimes \cdots \otimes \boldsymbol{M}$. The offset vector $\vec{O}^N$ contains outcomes in which at least one atom was not optically pumped.

Correcting for SPAM errors means solving $\vec{P}_\psi^N$ out of a measured $\vec{P}^N$. We numerically estimate $\vec{P}_\psi^N$ and $\vec{O}^N$ for each dataset by minimizing a cost $C = \|p^N \boldsymbol{M}^N \vec{P}_\psi^N + (1-p^N)\vec{O}^N - \vec{P}^N\|$ with the SciPy implementation of the SQSLP algorithm [54]. To ensure a physically valid outcome of $\vec{P}_\psi^N$, we impose boundary conditions $\sum_{i=1}^{2^N} \vec{P}_{\psi,i}^N = 1$ and $\sum_{i=1}^{2^N} \vec{O}_i^N = 1$. In Extended Data Fig.4, the raw and the SPAM-corrected results are presented for each dataset.

**Data Analysis.** We analyse the experimental data by reporting the Rydberg excitation probability $\langle n_i^z \rangle$ per region of interest (ROI) and by calculating the structure factor:

$$S(k_x, k_y) = \frac{1}{N} \sum_{i,j}^{N} \langle n_i^z n_j^z \rangle e^{ik_x(x_i-x_j)+ik_y(y_i-y_j)},$$

where we evaluate $k_x, k_y$ from $-3\pi/a$ to $3\pi/a$ in 129 steps. In the analysis, we post-select only the experimental runs that had a defect-free detection in the first image and exclude datapoints where the UV laser was delocked. A full overview of the runs included in each measurement is presented in Extended Data Table 1.

## B. UV laser system

**UV laser.** Two infrared fiber laser seeds (NKT Koheras BASIK X15 and NKT Koheras BASIK Y10) form the source of the light. After amplification with fiber amplifiers (NKT Koheras Boostik HP Y10 and Cybel Stingray BT-1550) and mode-matching the beams into a PPLN crystal (Covesion MSFG637-0.5-40), 633 nm light is generated by sum-frequency generation. Some leakage power is split into a path for a wavelength meter (HighFinesse WS8-30) and a path for frequency stabilization, see below. The main path of the 633-nm light is coupled into a low-finesse bow-tie cavity with a BBO crystal (Nortus Optronic GmbH) and then frequency doubled to 317 nm. After the light has passed the switching AOM and beam shaping optics, we estimate about 26 mW is available at the atoms. The measured Gaussian beam waist at the location of the atoms is about 280(5) μm in the horizontal direction (atom plane) and about 9.0(2) μm in the vertical direction (direction of tweezer propagation). With the transition dipole matrix elements in Ref. [55], we estimate a maximal Rabi frequency of $\Omega_0 = 2\pi \times 2.37$ MHz. This matches well the observed maximal Rabi frequency.

**Frequency stabilization.** The frequency of the UV laser is stabilized by locking the 633-nm light to a high-finesse (F ≈ 50000) cavity. Leakage light from the main setup is sent to the cavity and used for a Pound-Drever-Hall (PDH) lock. To provide absolute frequency stability, the cavity length is referenced to 689-nm light from our main laser that is spectroscopy locked to $^{88}$Sr atoms. The locking performance of the 633-nm light is measured by creating a beat note between the cavity transmission and a frequency-shifted path from the incoming light. We extract a single-sided power spectral density (PSD) by only considering the values at frequencies higher than the shifted frequency and doubling the amplitudes. Furthermore, we multiply the amplitudes by a factor of 4 to take into account the expected effect of frequency doubling and convert the spectrum to a laser frequency noise PSD by multiplying by the square of the frequencies, see Extended Data Fig.6b.

**Intensity noise.** Long-term intensity stability is measured with a monitor photodiode. We record 20-s long time traces and calculate the relative intensity noise (RIN) from the measured photodiode voltage V(t) as RIN(t)=(V(t)–$V_{avg}$)/$V_{avg}$, with $V_{avg}$ being the time-averaged voltage. Since for all physical purposes we care more about the Rabi frequency noise than the intensity noise and $\Omega \propto \sqrt{I}$, we define a noise parameter, $\epsilon(t)$ = RIN/2. Using Welch's algorithm [56, 57], the time-trace is chopped into 8-ms long windows that are half-overlapping, and an average PSD is obtained, see Extended Data Fig.6c.

**Pulse shape.** In experiments, the UV light is only sent onto the atoms during a few microseconds. The rise time of electronics and the AOM distort the shape of pulses, leading to deviations from the programmed ramp. We calibrate this effect by creating many pulses of variable nominal duration and recording their intensity on the photodiode. An example of a nominally 4-μs long pulse is presented as the blue trace in Extended Data Fig.6d. We confirm that the pulse shape does not lead to a large deviation from the intended pulse area, by calculating the integral of the square root intensity and plotting it over the nominal pulse duration in Extended Data Fig.6e.

Another effect that leads to intensity variation during the ramps performed in the experiment is the frequency dependence of the switching AOM diffraction efficiency. At constant RF power, we vary the frequency of the RF signal sent to the AOM and measure the power of the diffracted light with a photodiode. The powers are normalized with the power measured at 196.3 MHz — the AOM frequency at which we observe resonance, see below — and we take the square root to calculate the measured normalized Rabi frequency. The results are plotted in Extended Data Fig.6f.

In Extended Data Fig.6g, we compare the programmed (red) Rabi frequency of the pulse used to prepare the $\langle n_\uparrow \rangle$=6 antiferromagnetic state to the measured normalized Rabi frequency during the pulse (blue). The pulse duration is 7 μs, in which the AOM frequency is swept from 190.3 MHz to 220.3 MHz. Both the pulse distortion and the frequency dependence are visible in the blue trace. We visualize the effect of the non-constant Rabi frequency during the ramp by drawing the trajectory in the Δ,V-plane in Extended Data Fig.6h. The dotted line represents the programmed trajectory, and the solid line represents the one based on the measured Ω(t) in Extended Data Fig.6g. Time is depicted with vertical ticks in the trajectories. Each tick represents a duration of 300 ns. At early times, when Ω is ramped on, the trajectories are linear with constant V/$\Delta_i$, with $\Delta_i$ the initial detuning. During the frequency ramp, the programmed line stays horizontal because Ω is assumed to be constant. The measured trajectory also changes vertically, similar to the inverse of the pulse shape in Extended Data Fig.6g. At late times, as Ω is ramped off, both trajectories are straight lines with constant V/$\Delta_f$, with $\Delta_f$ the final detuning.

### C. Calibration of Hamiltonian parameters

**Resonance.** We perform Rabi spectroscopy to find the resonance of the 317-nm light with the 5s5p $^3P_0$ ↔ 5s61s $^3S_1$ transition in $^{88}$Sr. Experiments are performed at zero magnetic field. Atoms are rearranged into a 3 × 3 array with large spacing (≈18 μm) and imaged. Then, the atoms are incoherently pumped to 5s5p $^3P_0$ and illuminated with a 220 ns (π-)pulse of 317-nm light at different frequencies. At each point of the scan, the measurement procedure is repeated 100 times. We plot an example for one tweezer in Extended Data Fig.7a, see SI for fit details. The detuning is defined as the angular frequency shift from resonance, Δ=2π(f −$f_0$). Averaging over all tweezers, we measure the resonance condition at an AOM frequency shift of $f_0$=392.72(5) MHz. The absolute frequency is determined with a wavelength meter to within 30 MHz and is consistent with theoretical predictions [55].

**Single-atom Rabi oscillations.** We record single-atom Rabi oscillations to characterize the atom-light coherence. Atoms are loaded into the same 3 × 3 array as mentioned above and illuminated with resonant 317-nm light for a varying duration. Averaging 60 realizations, the probability of detecting an atom before and after the UV excitation is plotted in Extended Data Fig.7b. Error bars are the standard deviation of the mean. Simulations of expected decoherence due to the noise profiles of the laser intensity noise and laser frequency noise, following the methods highlighted in references [55, 57, 58] do not explain the decoherence. Rather, we estimate from calculations that the observed decoherence is dominated by shot-to-shot fluctuations, mostly due to alignment in the direction where the beam is smallest. We consider three types of shot-to-shot fluctuations: DC laser intensity noise, Doppler shifts, and misalignment. By combining the three noise sources and assuming them independent, a close match is observed, indicating that shot-to-shot noise limits the observed coherence. See SI for details of the fit function and exact parameters.

**Interaction strength.** The interaction strength between two Rydberg atoms is calibrated with quasi-adiabatic sweeps. There are two phase transitions in the two-atom system, which can be located in the classical regime of $\Omega \rightarrow 0$. For $\Delta<0$, the ground state has both atoms down: $|\downarrow\downarrow\rangle$. At $\Delta=0$, this state becomes degenerate with the $|\uparrow\downarrow\rangle$ and $|\downarrow\uparrow\rangle$ states. At $\Delta=V$, these become degenerate with the $|\uparrow\uparrow\rangle$ state. By probing the location of the upper resonance, we characterize the interaction strength for a given separation between the atoms. For five different separations, a tweezer pattern is created with six pairs of atoms, as displayed in Extended Data Fig.7c. The distances are discretized in Fourier units $d_{FFT}$, as described in Ref. [50]. In our system, $d_{FFT} \approx 0.45$ µm. After sorting atoms into isolated pairs, a picture is taken. The atoms are pumped to 5s5p$^3$P$_0$, and the Rabi frequency and detuning are ramped with the AOM during a few microseconds (Extended Data Fig.7d). The detuning is ramped linearly from far red-detuned to a specific value $\Delta_f$, which is varied. For the two atoms in a pair at $d=12d_{FFT}$, the detection probability is presented as a function of $\Delta_f$ in Extended Data Fig.7e. The data is post-selected on having both atoms present in the initial image. The solid lines are fits with a heuristically chosen double logistic function, see SI.

Repeating this experiment for different distances $d$, we obtain the data plotted in Extended Data Fig.7f. The solid line is a fit with V(d) =$C_6/d^6$. The data follows a $1/d^6$ scaling well. The fitted value for the coefficient is $C_6=2\pi \times 2.96 \times 10^7$ MHz $d_{FFT}^6$. Using our definition of $d_{FFT} \approx 0.45$ µm, this corresponds to $C_6 \approx 2\pi \times 2.5 \times 10^5$ MHz µm$^6$. We note that because of the uncertainty in the value of $d_{FFT}$, the $C_6$ value should not be taken as a precise determination.

**Benchmark at low $V$.** To calibrate the performance of our system, we prepare the antiferromagnetic ground state at low interaction strength of $V/\Omega=4.4$ by increasing the spacing between atoms in the Sierpiński gasket, see Extended Data Fig.8. Effectively, this yields no long-range, but only NN interaction. In this regime, the required detuning ramp covers a smaller frequency span due to the smaller spacing between intermediate states, allowing for a relatively slower ramp in $\Delta$ compared to the energy gaps between intermediate states. The lower interaction strength prevents the preparation of the intermediate states shown in the main text due to the lack of plateaus in a 1D trace of the phase diagram, but the larger spacing of the atoms reduces the effects of deformations in the atom positions, see the next section. These differences in the ramp, interaction, and pattern used lead to an excellent agreement with theory, where the remaining discrepancies are mainly related to SPAM errors.

**Trap deformations at tight spacings**
In the measurements presented in the main text, the exact outcome depends strongly on the exact tweezer geometry. Even a slight asymmetry can lead to a difference in interaction strength and result in a preferred site for the excitation. We suspect this is one of the major causes for the asymmetric magnetization in the measurements for the $\langle n_\uparrow \rangle$ =4,5 states. Here, we briefly summarize some possible causes for such an asymmetry, but more information and tests supporting our hypothesis can be found in the SI. Both possible causes are related to the weighted Gerchberg-Saxton (WGS) algorithm used to calculate the phase pattern for the SLM [59,60]. The first is the deformation of individual tweezer traps due to optical interference. Ideally, tweezers are assumed to have a cylindrically symmetric Gaussian shape, but in practice, optical aberrations and interference lead to deformations, especially at spacings proportional to the diameter of the tweezers. In the WGS algorithm, the optical phase is left as a free parameter to generate the desired intensity distribution, which can lead to unwanted interference between traps. Additionally, this can lead to interference effects outside of the focal plane that effectively skew the trapping potentials in the out-of-plane dimension [61].

A second effect influencing the tweezer geometry is rounding errors in the positions of the traps. When using the WGS algorithm, there is a trade-off between calculation efficiency and the discretization of grid points. The finite size of the SLM (N × N pixels) discretizes the number of programmable tweezer positions in the WGS algorithm. The hexagonal lattice underlying the investigated fractal has coordinates that are multiples of $\sqrt{3}$ when converted to Cartesian coordinates. This results in non-integer coordinates of the target tweezer positions on the square grid of the SLM. Since the WGS algorithm can only calculate integer coordinates, this leads to rounding errors that degrade the threefold symmetry of the pattern due to varying interaction strengths.

**Data availability**
The data are available from the corresponding authors on reasonable request.

**Code availability**
The codes are available at https://public.yoda.uu.nl/science/UU01/I0BH9I.html .

**Acknowledgements:** We are grateful to T. Macrí for fruitful discussions and for participating in the early stage of this research. We also acknowledge discussions with M. Aidelsburger. RCV, FS and CMS acknowledge financial support from the Netherlands Organization for Scientific Research (NWO, Grant No. 680.92.18.05). TSES and VZ acknowledge financial support from the Brazilian agency Coordenação de Aperfeiçoamento de Pesquisa de Pessoal de Nível Superior (CAPES) under the Netherlands Universities Foundation for International Cooperation (NUFFIC) exchange program (process number 88887.649143/2021-00). VZ acknowledges financial support from PNRR MUR Project No. PE0000023-NQSTI. We thank the NICIS Centre for High-Performance Computing, South Africa, for providing computational resources. CMS acknowledges the research program "Materials for the Quantum Age" (QuMat) for financial support. This program (registration number 024.005.006) is part of the Gravitation program financed by the Dutch Ministry of Education, Culture and Science (OCW). IHAK, YCT, AU, FS and RJCS acknowledge support from the Dutch National Growth Fund (NGF), as part of the Quantum Delta NL programme. The work has also received funding under Horizon Europe programme HORIZON-CL4-2021-DIGITAL-EMERGING-01-30 via project 101070144 (EuRyQa). We thank QDNL and NWO for grant NGF.1623.23.025 ("Qudits in theory and experiment").

**Author contributions** RCV performed exact diagonalization calculations for the first generation Sierpiński gasket and developed SIM-GRAPH, with which he calculated the phase diagrams and spin configurations for the second and third generations. He also wrote most of the theoretical part of the Methods and SI. TSES did the first calculations when the project started using exact diagonalization and variational mean field. He also investigated the correlations and calculated the phase diagrams for the spin susceptibility and entanglement entropy. VZ did numerical calculations using VMF and QMC. IHAK, AU, and YCT performed the experiments under the supervision of RJCS and FS. AU and IHAK wrote the experimental part of the Methods and SI. CMS conceived and led the project, supervised RCV, and wrote the paper with input from all authors.

**Competing interests** The authors declare no competing interests.

**EXTENDED DATA**

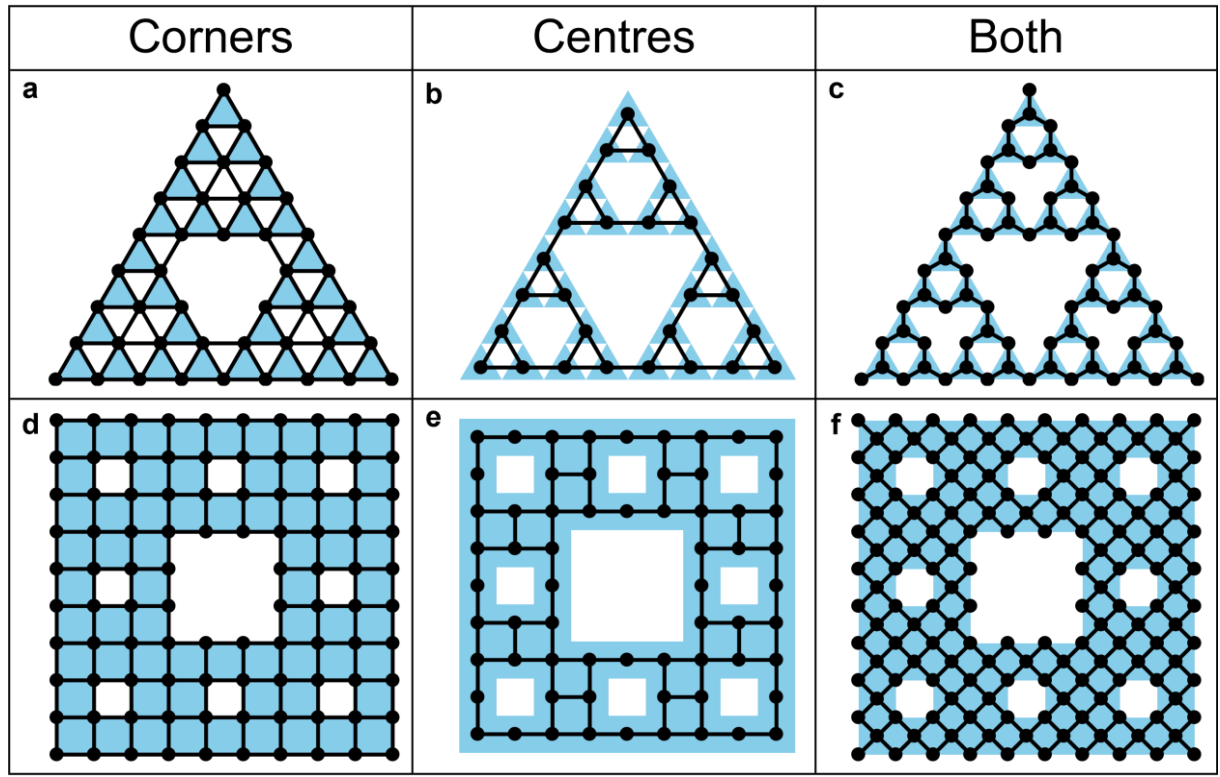

**Extended Data Fig. 1| Three ways of defining a lattice on a fractal. a-c**, Possible lattice definitions exemplified for the third-generation Sierpiński gasket; **d-f**, same for the second-generation Sierpiński carpet. For each unit cell, lattice points can either be placed at the corners (a and d), centres (b and e), or both (c and f). Then, NN connections are made using black-solid lines.

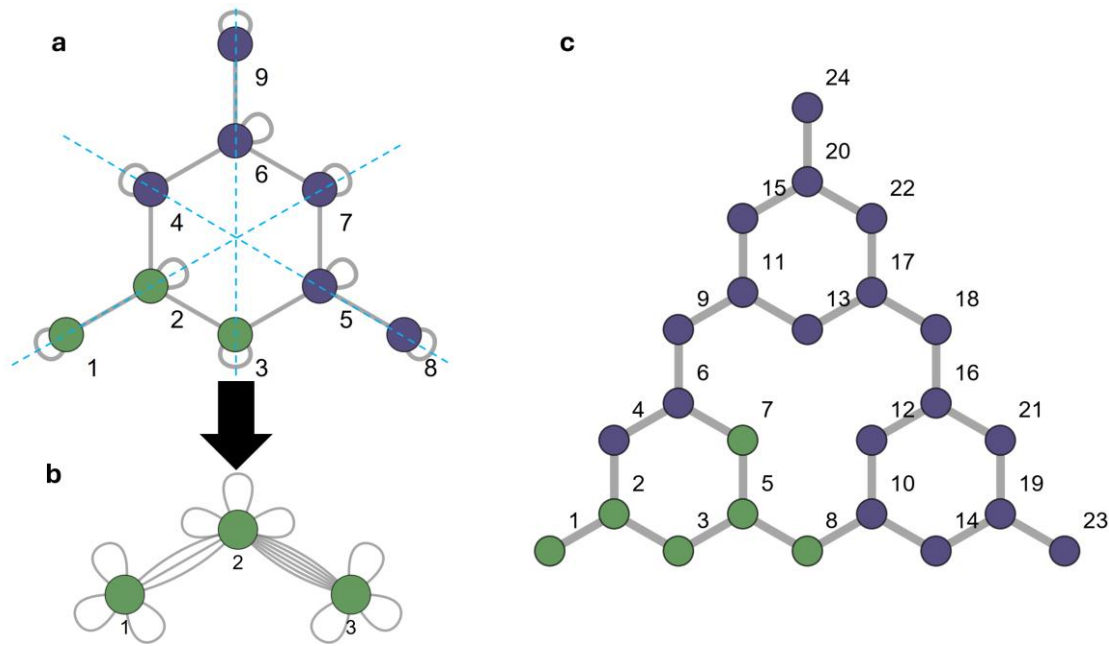

**Extended Data Fig. 2| Examples of SIM-GRAPH reduction. a**, First-generation Sierpiński gasket. The symmetry axes are shown in dashed-blue lines and the symmetric subset is highlighted in green. Interactions are denoted by grey lines. **b**, The connections of **a** are mapped onto their respective symmetry points, forming the reduced graph consisting of 3 instead of 9 sites. Repeated connections can be simplified by adding up their weights into a single stronger edge. **c**, The symmetric subset is highlighted for the second generation Sierpiński gasket, allowing for a reduction from 24 to 6 sites. For visual simplicity, self-interacting loops are not shown.

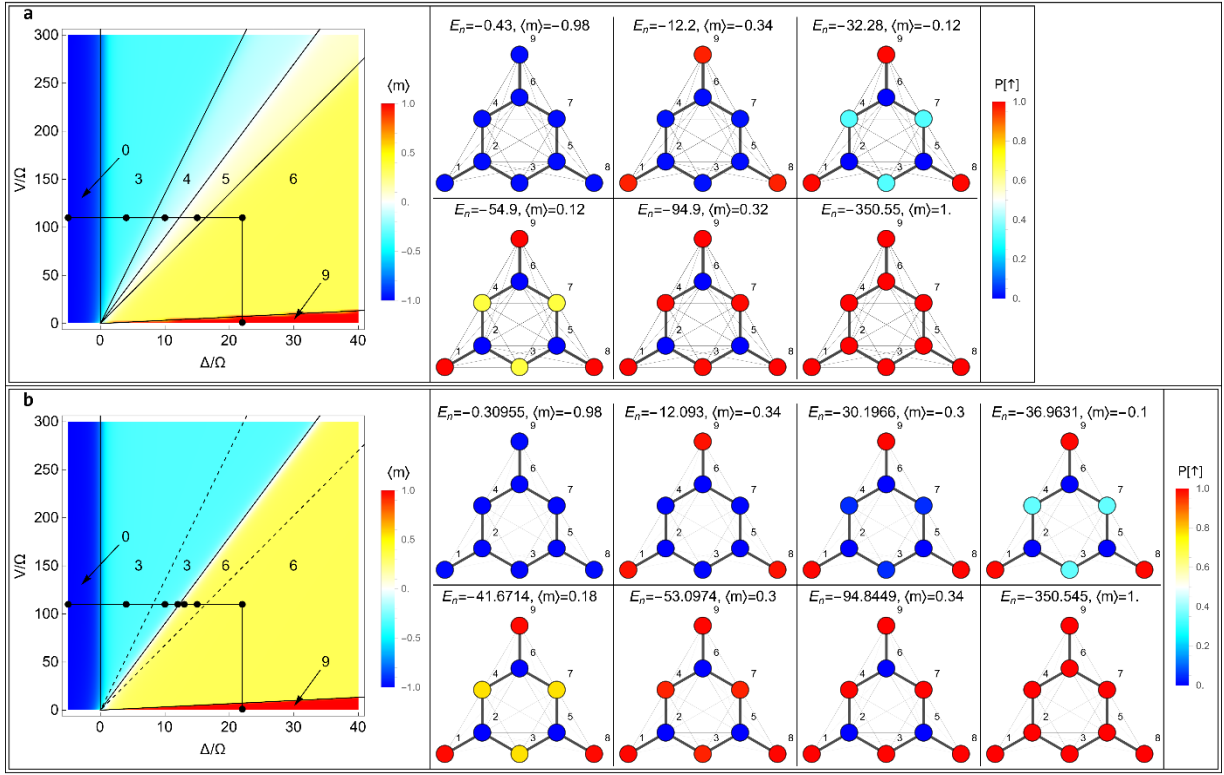

**Extended Data Fig. 3| The ground state magnetization phase diagram for the first generation hexagonal Sierpiński gasket using a, ED; b, SIM-GRAPH.** Within each domain of the phase diagram, we indicate the number of spins up. The phase transitions are highlighted using black lines and the states calculated at the black dots are presented on the right-hand side. The two intermediate phases, corresponding to four and five spins up are not correctly identified by SIM-GRAPH (see regions between the dashed lines in **b**. Instead, these phases are compressed into the narrow white line, as indicated by the snapshots of the spin configurations in **b** for ⟨m⟩ = −0.1 and ⟨m⟩ = 0.18.

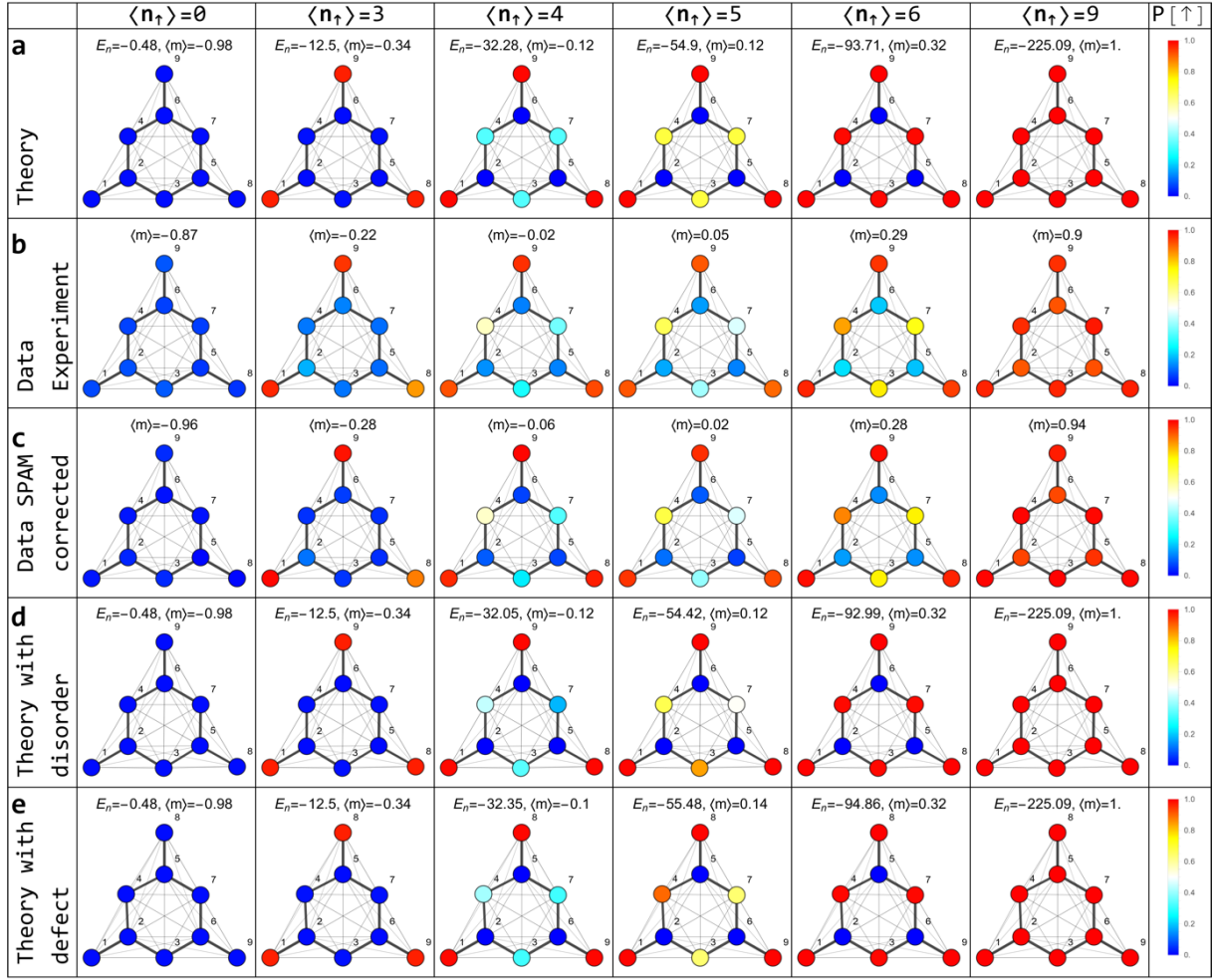

**Extended Data Fig. 4| Comparison of theoretical and experimental results including defects and disorder.** The same results as shown in Fig. 2 in the main text, but here we use an extreme colour-bar to highlight the differences. **a**, Theoretical results for the perfect lattice. **b**, Raw experimental data. **c**, SPAM corrected experimental data. **d**, Theoretical results with 1% Gaussian disorder on the position of the sites. **e**, Theoretical results for a defect on site 4, which is radially displaced outwards by 10% of the NN distance.

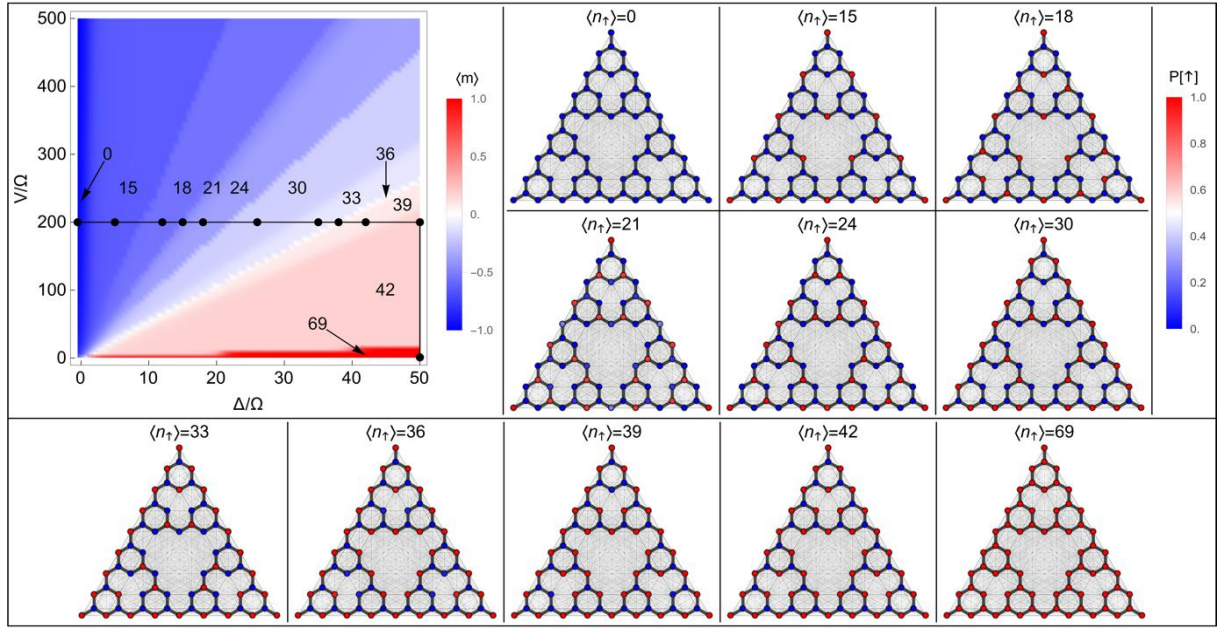

**Extended Data Fig. 5| Phase diagram and spin configurations for the third-generation Sierpiński gasket computed using SIM-GRAPH.** The spin configurations are evaluated at the black dots shown in the phase diagram, which are selected near the middle of each regime. The total number of spins increases by the symmetry factor (3) from $\langle n_\uparrow \rangle$=15 up to to $\langle n_\uparrow \rangle$=42, outside of which we find the standard ferro- ($\langle n_\uparrow \rangle = 69$) and antiferromagnetic ($\langle n_\uparrow \rangle = 42$) patterns. Notice the narrow regimes of $\langle n_\uparrow \rangle$=21 and $\langle n_\uparrow \rangle$=36. The regime of $\langle n_\uparrow \rangle = 27$ is not discernible. These results reiterate the ones obtained for the first- and second-generation Sierpiński gasket.

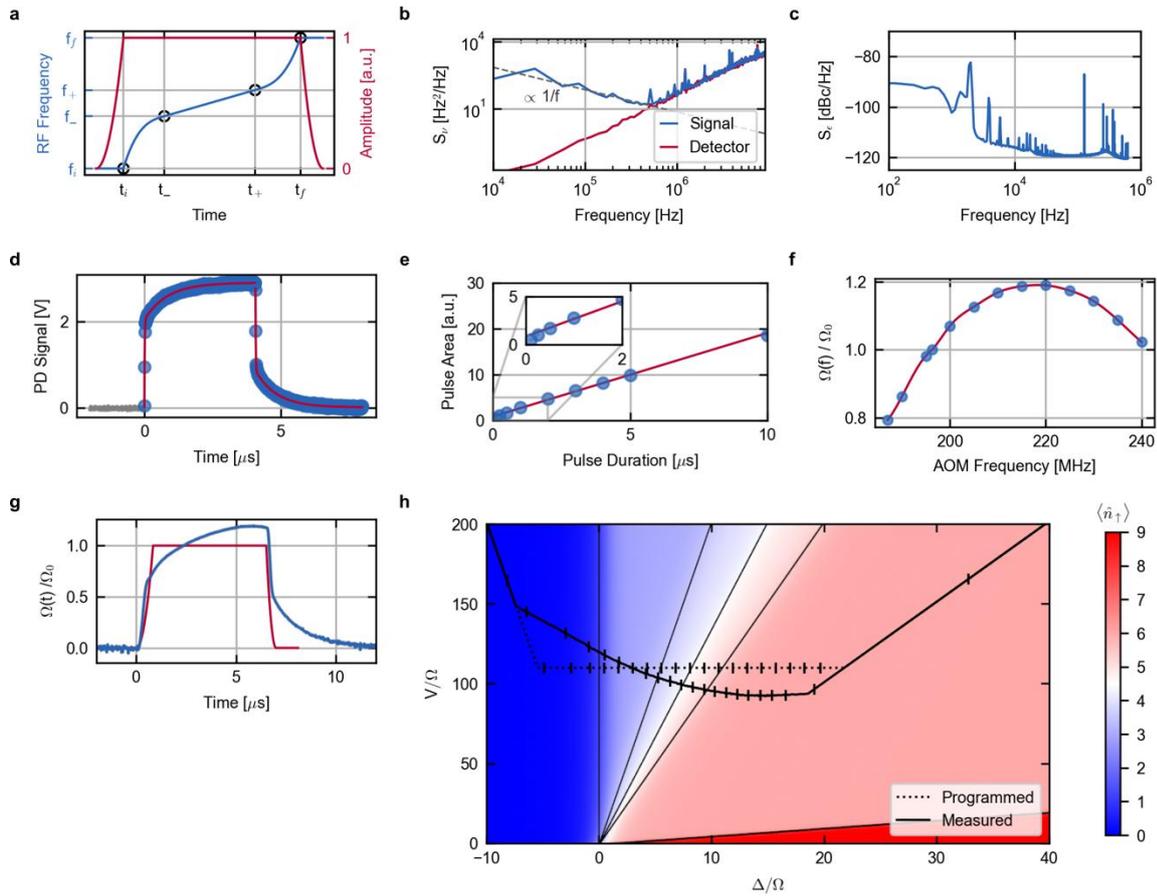

**Extended Data Fig.6| UV ramps, laser noise and pulses. a,** Schematic representation of the programmed UV ramps. At the start of the ramp, the RF amplitude (red) is ramped up quadratically to a nominal value and then kept constant until the frequency ramp (blue) is finished. Afterwards, the RF amplitude is ramped down again quadratically. The frequency ramp consists of three parts: it grows following an n-th order polynomial up until t−, then continues growing linearly until t+, after which it again grows polynomially. Here we draw n = 3, as used in the experiments. **b,** Laser frequency noise PSD (blue) obtained with the method described in the text. For frequencies > 500 kHz, the detector white noise (red) is the dominant contribution. **c,** Electric field amplitude noise power spectral density (PSD) of the UV laser measured with a photodiode during a 20-s long exposure. **d,** On shorter pulses used in the experiment, the beam shape is distorted as seen in the PD signal (blue) during a nominally 4-μs long rectangular pulse. Red traces are fits of a sum of two exponentials, with time constants of about 100 ns and 0.8 μs, which vary slightly with the pulse duration. **e,** The pulse area rises approximately linearly with the nominal pulse duration, as expected for rectangular pulses and despite the pulse shape distortion. See text for details. **f,** The expected Rabi frequency as a function of the switching AOM frequency, normalized by the Rabi frequency at resonance $\Omega$. The red trace is a cubic spline between the measured points. See text for details. **g,** The pulse distortion in **d** and AOM-frequency dependence of $\Omega$ in **f** lead to a non-constant Rabi frequency $\Omega(t)$ during the frequency ramp. The red line shows the programmed $\Omega(t)$ for the pulse used to prepare the $\langle n_\uparrow \rangle$ = 6 AFM state. The blue line is the observed Rabi frequency during the pulse. **h,** Parametric plot of the programmed (dotted) and measured (solid) trajectories in the Δ, V -plane when preparing the $\langle n_\uparrow \rangle$ = 6 state. Each vertical tick depicts a time step of 300 ns.

| | Frequency ramp | | | | Runs incl. | Frequent outcomes | | | | SPAM corrected outcomes | | | |
|---|---|---|---|---|---|---|---|---|---|---|---|---|---|
| $\langle n_\uparrow \rangle$ | $(t_i, \delta_i)$ | $(t_-, \delta_-)$ | $(t_+, \delta_+)$ | $(t_f, \delta_f)$ | | | | | | | | | |
| 0 | (-, −18.6) | | | (-, −18.6) | 2249 | : 56.6% | : 3.9% | : 4.6% | : 3.9% | : 86.3% | : 2.2% | : 2.9% | : 1.6% |
| 3 | (1.9, −12.0) | (4.5, 0.6) | (6.0, 6.0) | (6.6, 9.0) | 3892 | : 38.3% | : 7.2% | : 7.9% | : 4.8% | : 57.6% | : 7.7% | : 11.0% | : 4.2% |
| 4 | (2.0, −12.0) | (3.0, 0.7) | (7.0, 17.9) | (7.5, 22.0) | 3218 | : 20.5% | : 7.0% | : 8.8% | : 6.2% | : 29.5% | : 8.6% | : 12.4% | : 8.2% |
| 5 | (1.9, −12.0) | (2.7, 0.7) | (6.4, 30.0) | (6.8, 33.0) | 2428 | : 13.1% | : 10.2% | : 11.9% | : 4.9% | : 17.3% | : 14.3% | : 15.8% | : 6.3% |
| 6 | (2.0, −12.0) | (3.0, 0.7) | (7.3, 41.3) | (7.7, 48.0) | 3428 | : 26.8% | : 7.1% | : 7.8% | : 5.3% | : 34.1% | : 9.7% | : 10.5% | : 5.1% |
| 9 | (2.0, −12.0) | (3.0, 0.7) | (7.3, 41.3) | (7.7, 48.0) | 2348 | : 65.7% | : 5.4% | : 5.5% | : 5.3% | : 78.4% | : 5.8% | : 5.9% | : 4.8% |

**Extended Data Table 1| Details on the experimental data.** The datasets are labelled by the target number of excitations ⟨n↑⟩ at the end of the ramp. For each dataset, we denote the four programmed coordinate pairs (t in μs, δ in MHz, where δ=f-392.6 MHz is the detuning from resonance, see Extended Data Fig.6a) that determine the RF frequency sweep shape. We also present the number of experimental realizations in the datasets. In the last two columns, we show the four most frequently observed outcomes and the four most frequently inferred outcomes after SPAM correction per dataset, respectively, with figures in which a red dot indicates an atom was detected and an empty circle means a vacancy.

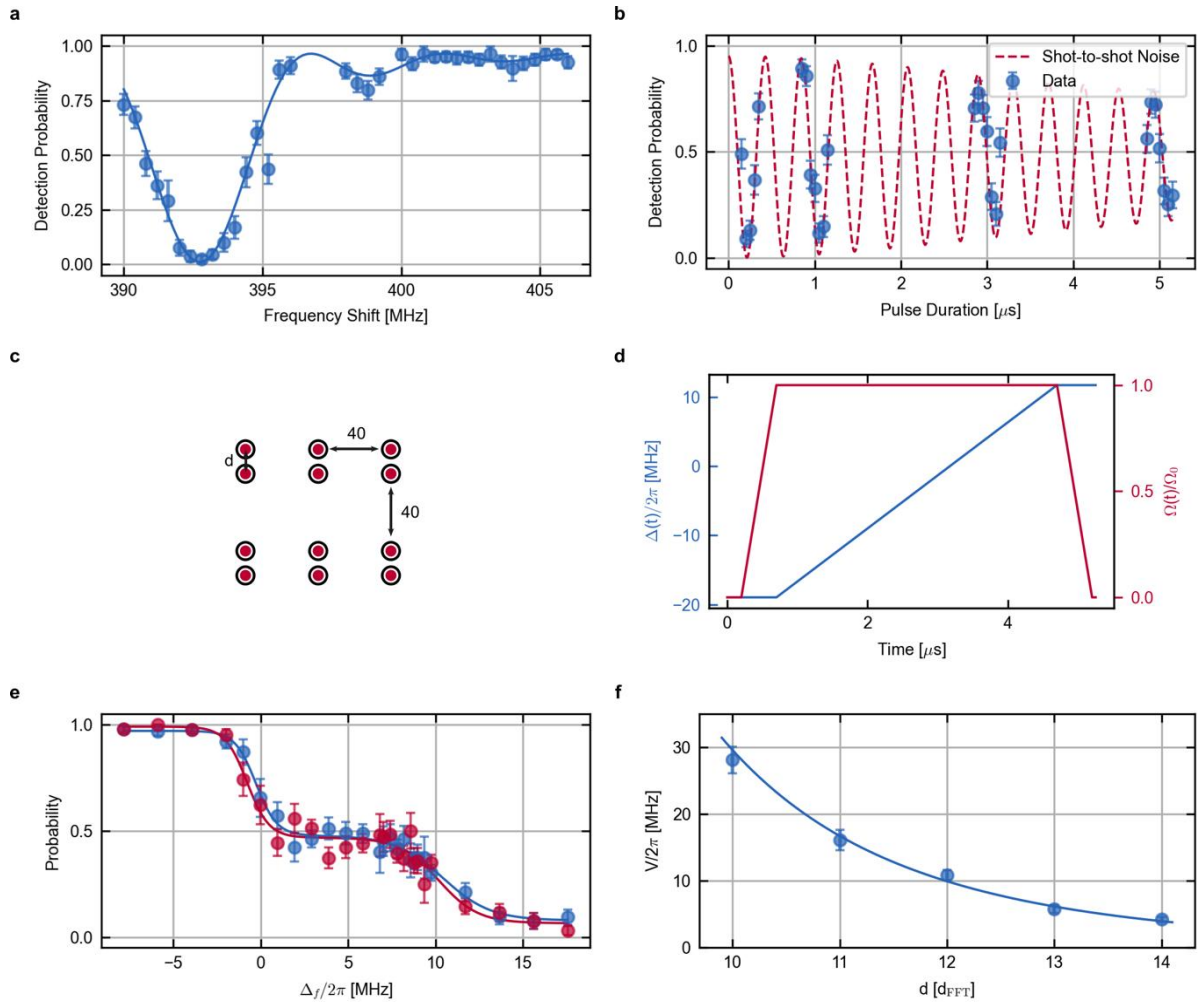

**Extended Data Fig.7| Calibration of Hamiltonian parameters. a,** Rabi spectroscopy of the 5s5p $^3P_0$ ↔ 5s61s $^3S_1$ transition for a single tweezer. The pulse duration was 200 ns. The data points are detection probabilities over 100 repetitions. Error bars are standard deviations of the mean. **b,** Rabi oscillations compared to shot-to-shot noise model. **c,** Tweezer patterns used for the interaction calibration scan. Six pairs of atoms are isolated by $d$ Fourier units (see text). Each pair is separated by a large distance (40 $d_{FFT}$) to isolate the two-atom systems. **d,** Example ramp profile of the Rabi frequency (red) and detuning (blue). **e,** The detection probability of the atoms depends on the final detuning Δf. Red and blue traces are data for the two atoms in a single pair, separated by $d$ = 12 $d_{FFT}$. By fitting two double logistic functions (solid lines, see SI for equation), we obtain locations of the resonances. The interaction strength is extracted as the location of the upper resonance. **f,** Obtained interaction for 5 different pair separations. The solid line is a fit of the interaction strength.

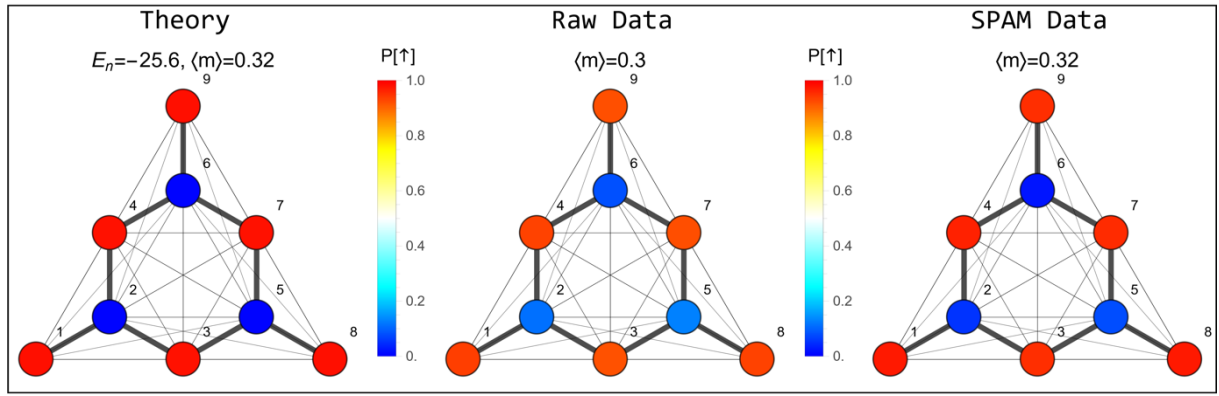

**Extended Data Fig.8| Benchmark of adiabatic state preparation at low V.** The antiferromagnetic state is compared between theory and experiment for a weaker interaction strength, namely at $V/\Omega = 4.44$ and $\Delta/\Omega = 4.44$. We find excellent agreement in this regime. The theory predicts 0% and 98% probability for the two sub-lattices, whereas the measured probabilities are on average 10% and 93% for the raw data, and 4.8% and 96.4% for the SPAM corrected data.

# Supplementary Information for "Control of single spin-flips in a Rydberg atomic fractal"

R. C. Verstraten*, I. H. A. Knottnerus*, Y. C. Tseng, A. Urech, T. S. do Espirito Santo,

V. Zampronio, F. Schreck, R. J. C. Spreeuw, and C. Morais Smith

August 20, 2025

## Contents

## 1 Supplementary Equations: Adiabatic ramp function for AWG

Here we describe in detail the shape of the RF signal used for ramping the 317-nm light described in Methods. On the AWG, we program a voltage trace $V(t) = u(t)\sin(\varphi(t))$ that describes the RF signal that is sent, after amplification, to the AOM. The ramp consists of five stages, see Extended Data Fig.2a. In the first part of the ramp, we program a signal with constant frequency and quadratically increasing RF amplitude, shown as red trace. Next, we program a constant RF amplitude and vary the RF frequency (blue trace) in three steps: In the first part, $t_i < t \leq t_-$, the frequency increases polynomially from $f_i$ to $f_-$. For $t_- < t \leq t_+$, the frequency grows linearly from $f_-$ to $f_+$. When $t_+ < t \leq t_f$, the frequency again grows polynomially from $f_+$ to $f_f$. By imposing that both the frequency and the phase should be continuous at $(t_-, f_-)$ and $(t_+, f_+)$, we construct the following spline:

$$\varphi(t) = 2\pi \begin{cases} f_i t + A & t \leq t_i \\ f_- t + \frac{f_- - f_i + \frac{(f_+ - f_-)(t_i - t_-)}{t_+ - t_-}}{(n+1)(t_- - t_i)^n}(t_- - t)^{n+1} + \frac{1}{2}\frac{f_+ - f_-}{t_+ - t_-}(t - t_-)^2 + B & t_i < t \leq t_- \\ f_- t + \frac{1}{2}\frac{f_+ - f_-}{(t_+ - t_-)}(t - t_-)^2 + C & t_- < t \leq t_+ \\ f_+ t + \frac{f_f - f_+ - \frac{(f_+ - f_-)(t_f - t_+)}{t_+ - t_-}}{(n+1)(t_f - t_+)^n}(t - t_+)^{n+1} + \frac{1}{2}\frac{f_+ - f_-}{t_+ - t_-}(t - t_+)^2 + D & t_+ < t \leq t_f \\ f_f t + E & t > t_f \end{cases} \quad \text{(S1)}$$

with integration constants:

$$A = \frac{\varphi_0}{2\pi},$$
$$B = (f_i - f_-)t_i - \frac{1}{2}\frac{(f_+ - f_-)(t_i - t_-)^2}{t_+ - t_-} - \frac{f_- - f_i + \frac{(f_+ - f_-)(t_i - t_-)}{t_+ - t_-}}{n+1} + A,$$
$$C = B,$$
$$D = \frac{1}{2}\frac{f_- - f_+}{t_+ + t_-} + C,$$
$$E = (f_+ - f_f)t_f + \frac{1}{2}\frac{(f_+ - f_-)(t_f - t_+)^2}{t_+ - t_-} + \frac{f_f - f_+ - \frac{(f_+ - f_-)(t_f - t_+)}{t_+ - t_-}}{n+1} + D.$$

Here, $\varphi_0$ is an initial phase and $n$ denotes the order of the polynomial that is used. This shape of frequency ramp has the benefit of allowing a big gradient at early and late times, when the energy gap between the ground state and first excited state is large, but having a minimal linear gradient when the energy gap is small. When the frequency ramp is finished, the amplitude is ramped down quadratically. In experiments, we used $n = 3$ and ramped the frequency over $6\,\mu$s or $7\,\mu$s, depending on the final frequency value. The values for the frequencies $f_i, f_-, f_+, f_f$ varied per ramp in the experiment and are presented as detunings (e.g. $\delta_i = 2 \times (f_i - f_0)$, with $f_0$ the AOM frequency at resonance and the factor 2 because of the double-pass configuration) in Extended Data Table 1. It should be noted that experimental errors distort the pulse shape, such that the atoms experience a different intensity than programmed, as mentioned in the Methods.

# 2 Supplementary Notes: Additional data analysis and statistics

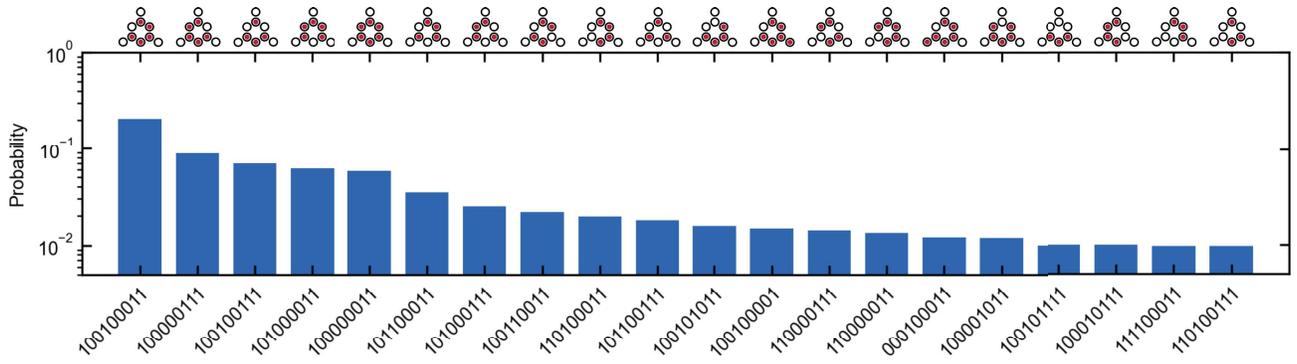

**Supplementary Figure** 1: Histogram of the 20 most frequently observed bitstrings in the data for the $\langle n_\uparrow \rangle = 4$ state. On the top axis a graphical depiction is presented of each bitstring. Filled circles denote that an atom was detected in an ROI, while empty circles denote the opposite. In 35.5 % of the realizations, one of the three expected states, with all corners and one edge excited, is measured. In total, $2^9 = 512$ different outcomes are possible.

Although in Fig. 2 of the main text only the averaged excitation probability is presented, another interesting way of analyzing the data is to look at the probability of measuring a specific outcome. This way, one gains insight in the distribution of outcomes. For each realization, we record the outcome as a bitstring, where we label $|\downarrow\rangle = 0$ and $|\uparrow\rangle = 1$. For example, "000000000" denotes all atoms were detected and thus assumed to be in $5s5p\,^3P_0$. The probabilities of the 20 bitstrings that occur most frequently in the dataset for the $\langle n_\uparrow \rangle = 4$ data are plotted in Supplementary Fig. 1. Only five bitstrings occur more frequently than 5 %. On the top axis, a graphical depiction is presented for all bitstrings. In the depictions, a red dot denotes the presence of an atom (i.e., "0") and an unfilled circle denotes the absence of an atom (i.e., "1"). Three of the five most frequent outcomes are expected for the $\langle n_\uparrow \rangle = 4$ state —three corners and one of the edges excited. The other two are a state with only the corners up and one with two excitations on the edges. Supplementary Fig. 1 shows that the asymmetry observed in the averaged plots of Fig. 2 of the main text stems largely from a preference of exciting one of the three expected bitstrings ("100100011": 20.5 %) over the other two ("100000111": 8.8 % and "101000011": 6.2 %). Here, we used the same numbering convention as in the main text. The total probability of reaching one of the three $\langle n_\uparrow \rangle = 4$ outcomes is 35.5 %. In Extended Data Tab.1, we present a visualization of the four most frequently observed outcomes for all datasets.

# 3 Supplementary Methods: Calibration of parameters in Hamiltonian

## 3.1 Determination of the resonance condition

For the determination of the resonance condition presented in Extended Data Fig. 2a, the solid line is a fit of

$$P_1(f) = A \left( 1 - \frac{\Omega^2}{\Omega^2 + \Delta^2} \sin^2 \left( \frac{1}{2} \sqrt{\Omega^2 + \Delta^2} t \right) \right), \tag{S2}$$

where $A$ is a constant to account for experimental losses, $\Omega$ is the Rabi frequency and $t = 220\,\text{ns}$ is the pulse duration. The detuning is defined as the angular frequency shift from resonance, $\Delta = 2\pi(f - f_0)$. Averaging over all tweezers, we measure the resonance condition at an AOM frequency shift of $f_0 = 392.72(5)\,\text{MHz}$.

Long-term effects, such as electric field and magnetic field drifts, can influence the exact position of the reference condition. We check the magnetic field at the atoms' location regularly with spectroscopy on the $5s^2\,^1S_0 \leftrightarrow 5s5p\,^3P_1$ ($m_J = \pm 1$) transition. This allows us to ensure that the magnetic field drifts stay within several $10\,\text{mG}$. This leads to a small uncertainty on the order of several $10\,\text{kHz}$. Our experiment does not have a direct way to measure and compensate for electric fields. In line with Ref. [1], a UV lamp (UVGO 365nm) is installed at the bottom of the science chamber, which can be used to illuminate the top window, inducing charge redistribution on the dielectric surface closest to the atoms (at a distance of $\approx 1\,\text{cm}$)), but we saw no clear improvement.

## 3.2 Single-atom Rabi oscillations

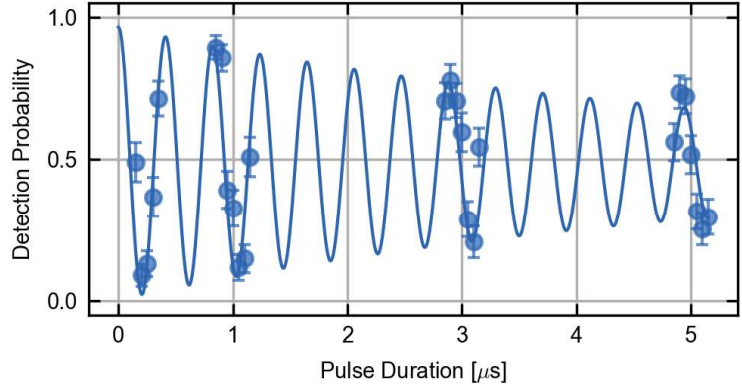

**Supplementary Figure** 2: Rabi oscillations of single atoms in a single tweezer. Data points are averages over 60 repetitions. Error bars are standard deviations of the mean. The solid line is a fit using Eqn. S3.

We record single-atom Rabi oscillations to characterize the atom-light coherence. Atoms are loaded into the same $3 \times 3$ array, as mentioned in Methods, and illuminated with resonant 317-nm light for a varying duration. Averaging over 60 realizations, the probability of detecting an atom before and after the UV excitation is plotted in Supplementary Fig. 2. Error bars are the standard deviation of the mean. The solid line is a fit using:

$$P_1(t) = A \left( \frac{1}{2} - \frac{e^{-t/\tau}}{2} \cos(\Omega t) \right). \tag{S3}$$

Here, $A$ is again a constant accounting for preparation losses and $\Omega$ is the Rabi frequency. The Rabi frequency averaged over all tweezers is $\Omega = 2\pi \times 2.39(3)\,\text{MHz}$. This is the maximum Rabi frequency we can achieve in experiments.

The contrast is assumed to decay exponentially, with a characteristic timescale $\tau \approx 5.5\,\mu\text{s}$ obtained from the fit. Simulations of expected decoherence due to the noise profiles of the laser intensity noise and laser frequency noise following the methods highlighted in references [2–4] do not explain the decoherence. Rather, we estimate from calculations that the observed decoherence is dominated by shot-to-shot fluctuations, mostly due to alignment in the direction where the beam is smallest. We consider three types of shot-to-shot fluctuations: DC laser intensity noise, Doppler shifts, and misalignment.

**DC laser intensity noise** is readily measured with a photodiode using a long exposure or by adding multiple smaller exposures [5]. We measure an RMS value of $\sigma_I = 0.6\,\%$ for the intensity, which translates to an RMS value of $\sigma_\epsilon = 0.3\,\%$ on $\Omega$. Alternatively, integrating in Extended Data Fig. 2b over all frequencies yields the same result. DC intensity noise results in a Gaussian decay envelope with characteristic time $\tau = \frac{\sqrt{2}}{\sigma_\epsilon \Omega} \approx 31\,\mu\text{s}$ [2, 5].

**Doppler shifts** arise from a non-zero momentum of atoms upon release before the UV excitation. Assuming a thermal distribution of motional levels, the expected momentum probability function is a Gaussian with standard deviation $\sigma_p$ [6, 7]

$$\sigma_p = \sqrt{\frac{\hbar m \omega}{2} \coth\left(\frac{\hbar \omega}{2 k_B T}\right)}. \tag{S4}$$

Here, $m$ is the atomic mass, $\omega$ is the trap frequency and $T$ is the temperature. We introduce a normalized frequency shift $\xi = \frac{2\pi p}{\lambda m \Omega}$ and consider $m = 88$ amu, $\omega \approx 2\pi \times 89$ kHz, and $T \approx 10\,\mu$K. To first order, the Doppler shift results in a decoherence envelope of $\left(1 + \Omega^2 t^2 \sigma_\xi^4\right)^{-1/4}$ [2]. With our numbers, we estimate the normalized frequency deviation to be $\sigma_\xi \approx 0.042$. The contrast is expected to be reduced to $1/e$ in about $280\,\mu$s.

**Misalignment** of the atoms with the 317-nm light happens due to two effects. First, upon release, the position of the atoms can be approximated to be normally distributed with a standard deviation:

$$\sigma_x = \sqrt{\frac{\hbar}{2m\omega} \coth\left(\frac{\hbar \omega}{2 k_B T}\right)}. \tag{S5}$$

Here, we consider the axial trapping frequency $\omega \approx 2\pi \times 13$ kHz, because that is the axis in which the UV beam shape is smallest. This effect is estimated to be $\sigma_{\text{atom}} \approx 0.38\,\mu$m. A second effect is drift of the UV beam position. We characterize this using a four-quadrant photodiode (Koheron 4QPD-100k) in the focus of the 317-nm light after the chamber and find the position distribution to be well-approximated by a Gaussisan with $\sigma_{\text{beam}} \approx 0.85\,\mu$m. Assuming both processes are independent, the variances are summed to obtain the total fluctuation. Normalizing by the waist $w$ to obtain a displacement $d = \delta x/w$ and following an analysis similar to Ref. [5] we obtain a decoherence envelope that scales as $\left(1 + 4\Omega^2 t^2 \sigma_d^4\right)^{-1/4}$. We estimate the normalized displacement deviation to be $\sigma_d \approx 0.1$ and the contrast to decay to $1/e$ in about $24\,\mu$s.

Combining the three noise sources and assuming them independent, the envelopes can be multiplied and we obtain a total expression for the expected detection probability:

$$P_1(t) \propto 1 - \left(\frac{1 - \sigma_\epsilon^2}{2} - \frac{e^{-t^2 \Omega^2 \sigma_\epsilon^2/2}}{\left(1 + \Omega^2 t^2 \sigma_\xi^4\right)^{1/4} \left(1 + 4\Omega^2 t^2 \sigma_d^4\right)^{1/4}} \frac{\cos(\Omega t + \varphi(t))}{2}\right). \tag{S6}$$

The phase accumulation

$$\varphi(t) \approx \arg\left((1 - 2i\Omega t \sigma_d^2)^{-1/2} \sigma_\xi (1 - i\Omega t \sigma_\xi^2)^{-1/2}\right) \tag{S7}$$

can be obtained by the full calculation of the integral of Eqn. (S2) expressed in terms of $\epsilon$, $\xi$, and $d$ times the three probability density functions. The full derivation is tedious, and $\varphi(t)$ remains small for the parameters presented here.

Extended Data Fig.3b compares Eqn. (S6) (red dashed line) with the Rabi oscillation data. A close match is observed, indicating that shot-to-shot noise limits the observed coherence. We note that for adiabatic preparation, the exact value of $\Omega$ often does not matter, as long as one fulfills the adiabaticity condition.

## 3.3 Calibration of the interaction strength

To fit the data presented in Extended Data Fig. 3d, we use a heuristically chosen double logistic function:

$$P_1(\Delta_f) = m_0 + \frac{m_1 - m_0}{1 + \exp\left(-\frac{\Delta_f - \Delta_{0 \to 1}}{\sigma_{0 \to 1}}\right)} + \frac{m_2 - m_1}{1 + \exp\left(-\frac{\Delta_f - \Delta_{1 \to 2}}{\sigma_{1 \to 2}}\right)}. \tag{S8}$$

The values $\Delta_{0 \to 1}$ and $\Delta_{1 \to 2}$ represent the resonances when the system goes from $0 \to 1$ and $1 \to 2$ respectively. The widths of the resonances are determined by $\sigma_{0 \to 1}$ and $\sigma_{1 \to 2}$. $m_0$, $m_1$, and $m_2$ are asymptotes that represent the survival at different magnetizations, after correction for losses in preparation and imaging. The interaction is determined as $V(d) = \Delta_{1 \to 2}$ from the fit.

# 4 Supplementary Methods: Characterization of experimental uncertainties

The result of the simulation is measured as the presence or absence of atoms in tweezers. In the experiment, there are many steps such as imaging or optical pumping that are not part of the physics we want to study, but do contribute to the outcome of the measurement. By calibrating the efficacy of the individual steps in separate experiments, we quantify the errors induced by each step. A model of how the relevant experimental steps influence our measurement outcomes is presented in Supplementary

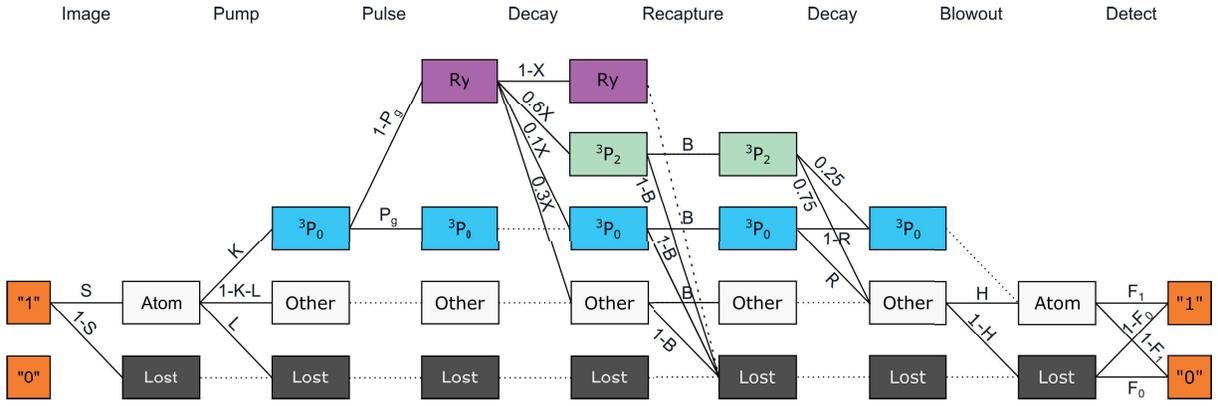

**Supplementary Figure** 3: Schematic depiction of the sources of error in our experimental sequence. At the left we start each run with post-selected runs that resulted in a "1" in the first image. Then we consider from left to right the different processes and how they change the atomic state. See main text for an explanation of each step. Dotted lines represent unit probability. Note that "1" and "0" denote the presence of an atom here, and not the excitation to a Rydberg state.

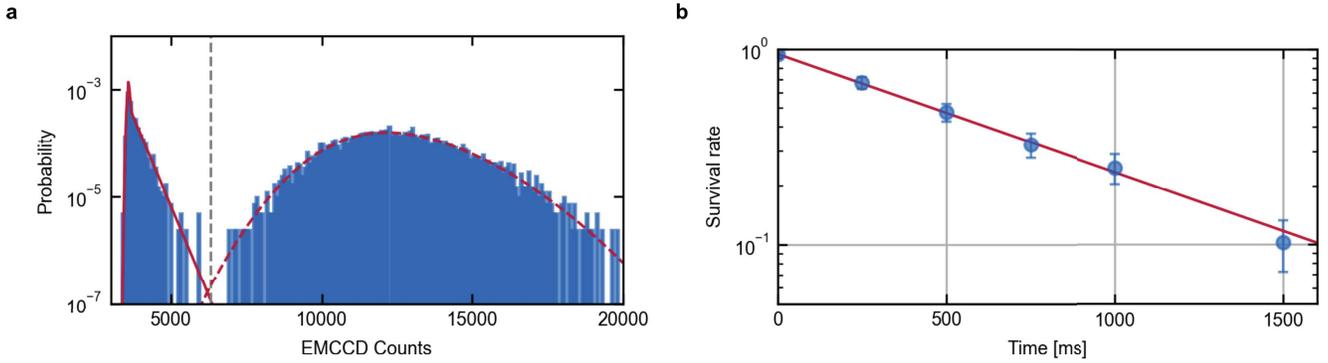

**Supplementary Figure** 4: **a,** Histogram of the EMCCD counts for one of the ROIs. Two peaks are visible, corresponding to zero or one atoms detected. Red solid and dashed lines are fits of the distributions of both peaks. The grey line is the binarization threshold above which the detection algorithm labels the signal as having an atom. **b,** Detection probability of an atom after being pumped to $5s5p\,^3P_0$, waiting for a variable amount of time, sending a blow-out pulse and repumping into the ground state. The red line is a fit of an exponential decay. A typical decay time constant of $\tau \approx 0.9\,\text{s}$ is observed. Data points are averaged over $100$ repetitions and error bars are the standard deviation of the mean.

Fig. 3. Following the image after rearrangement, the atoms undergo a series of steps depicted as lines that redistribute the populations in different states depicted by the blocks. Throughout the experiment different atomic states become relevant, and the labels on the blocks change. In the first part of this section, we explain and present characterization measurements of the individual steps in the diagram.

We note that our labeling convention is slightly different than that of *bright* and *dark* states, commonly used in literature [6, 8]. In our case, with long imaging durations and repumping light present throughout all images, we estimate all states to fluoresce (and thus be bright), including metastable states and Rydberg states. Therefore, we chose "Lost" as the only state in which an atom does not fluoresce in the images.

## 4.1 Measurement of individual efficacies

The general correction strategy is outlined in our earlier work [9] and is closely related to that of other groups [6]. All measurements in this work are based on two images. In the first image, the filling fraction after rearrangement is checked. For some experiments such as the simulation presented in the main text, the data is post-selected such that only runs with an atom detected in every desired tweezer are taken into account. The second image is taken after the experiment, which contains a step with an unknown probability of losing an atom. A raw measurement outcome is calculated as conditional probability $P(I_1|I_0)$ that an atom is detected in the second image ($I_1$), given that it was detected in the first image ($I_0$). By writing out this expression explicitly for each calibration experiment and solving for the unknown probability we obtain a characterization of the experimental step.

**Detection Fidelities** To characterize the fidelity of detecting an atom in a tweezer, single atoms are loaded stochastically in a $6 \times 6$ array and rearranged into the desired 9-atom geometry and an image is taken. To avoid having almost no signals corresponding to no atoms in tweezers, the rearrangement success for each atom is intentionally reduced to around $75\%$ for this particular measurement. After repeating the measurement 5000 times, a two-peaked histogram of the counts collected on the EMCCD camera is made for each ROI. By fitting the zero-atom and one-atom peaks in the histogram and comparing the areas under the fits to a preset binarization threshold, we obtain values for the true-positive ($F_1$) and true-negative ($F_0$) detection fidelities. An example histogram and its fits are presented in Supplementary Fig. 4. The grey dashed line depicts the threshold value, which is the same for all ROIs. Averaged over the nine ROIs, the fidelities are $F_1 = 0.9995(3)$ and $F_0 = 0.9998(2)$.

**Imaging Survival** The imaging survival rate is characterized by taking a second image after the first image and looking at the conditional probability $P(I_1|I_0)$ of detecting an atom in both images. This measurement is also repeated around 5000 times. Given the above detection fidelities and a probability $p$ of having an atom in a tweezer, the corrected image survival $S$ is given as [6, 9]:

$$S = \frac{(P(I_1|I_0) + F_0 - 1)(pF_1 + (1-p)(1-F_0))}{pF_1(F_1 + F_0 - 1)}, \tag{S9}$$

$$\approx \frac{P(I_1|I_0) + F_0 - 1}{F_1 + F_0 - 1}. \tag{S10}$$

In the second line, we made the assumption that $p \approx 1$, which is valid after rearrangement. Averaged over the array the imaging survival is $S = 0.9967(10)$.

**Blow-out survival** To blow out atoms in $5s^2\,^1S_0$, we shine $150\,\mu$W of 461-nm light in a 1.4-mm diameter beam for 2 ms onto the atoms while having 707-nm light present. The probability for atoms in $5s^2\,^1S_0$ to survive such a blow-out pulse is measured by sending such a pulse in between the two images. We define the pulse survival probability as $H$ and obtain an equation similar to Eqn. (S10):

$$H = \frac{P(I_1|I_0) + F_0 - 1}{S(F_1 + F_0 - 1)}. \tag{S11}$$

Repeating this experiment for more than 2000 times, we obtain an average survival of $P(I_1|I_0) = 0.0017(9)$. This leads to a corrected outcome of $H = 0.0013(10)$. We ignore the effect of the blow-out pulse on the atoms in the $5s5p\,^3P_0$ state, which is justified by the low off-resonant scattering rate of the 461-nm light on the relevant transitions for $5s5p\,^3P_0$.

**Loss of atoms in $5s5p\,^3P_0$** During the experiments, off-resonant scattering of the tweezer light and vacuum collisions limit the lifetime of the metastable $5s5p\,^3P_0$ state. We measure the decay time of the $5s5p\,^3P_0$ state by incoherently pumping atoms into the state, waiting a variable amount of time, and then sending a blow-out pulse before repumping the atoms back by shining 679-nm and 707-nm light onto them. The detected fraction of atoms decays exponentially with the wait time as plotted in Supplementary Fig. 4b. An exponential fit gives an estimated $1/e$-decay time of $\tau \approx 0.9$ s. The decay due to Raman scattering is only relevant when the tweezers are on and before the atoms are repumped, which is for approximately 5-ms, including the duration of the blow-out pulse. This results in an $5s5p\,^3P_0$ decay probability of $R \approx 0.006$.

**Optical pumping and losses** To prepare the atoms in the $5s5p\,^3P_0$ state, the atoms are illuminated for 30 ms with 689-nm, 688-nm, and 707-nm light. This incoherent pumping requires atoms to scatter many photons and induces heating and potentially atom loss. Following the example of reference [6], we distinguish three outcomes of the optical pumping: atoms can be successfully transferred with a probability $K$; atoms can be lost with probability $L$; or they remain in the traps, but not in $5s5p\,^3P_0$ with a probability $1 - K - L$. We calibrate $K$ and $L$ in two similar sequences. For the loss rate measurement, we apply the optical pumping, repump the atoms afterwards and image the atoms again. The corrected loss rate can then be found as:

$$L = 1 - \frac{P(I_1|I_0) + F_0 - 1}{S(F_1 + F_0 - 1)}. \tag{S12}$$

We measure a typical pumping loss of $L = 0.0050(22)$.

For the pumping success, an additional blow-out pulse is used to remove atoms that are still in the trap but not in $5s5p\,^3P_0$. We assume atoms that are lost due to Raman scattering of tweezers with probability $1 - R$ to have the full probability of being expelled by the blow-out pulse. Under this assumption, we can write the corrected pumping success rate as:

$$K = \frac{P(I_1|I_0) + F_0 - 1}{S(F_1 + F_0 - 1)(1-R)(1-H)} - \frac{H(1-L)}{(1-R)(1-H)}. \tag{S13}$$

In the limit $H \to 0$, this becomes similar to the result in reference [6], but with the factor $1 - R$ in the denominator. With our current definitions, we find this correct, as a higher decay rate $R$ leads to a lower raw detection probability and therefore should lead to higher corrected value. Averaged over the array, we measured a pumping success of $K = 0.989(3)$.

**Tweezer blink survival**   We measure the tweezer blink survival with a sequence similar to the one used for measuring the pumping, but now including an additional blink of $9.4\,\mu$s before the blow-out pulse. The $9.4\,\mu$s correspond to the duration of the blinks in most ramps of the experiment. Specifically for this experiment, we assume that the blink only affects atoms in $5s5p\,^3P_0$, because those atoms have been heated the most from the pumping process and therefore have the largest probability to be lost. This seems to contradict the decay channels in Supplementary Fig. 3, where all lowlying states are assumed to have the same blink survival probability $B$. We believe that assumption is valid for the experiments including UV light, because the majority of the atom population in the lowlying states has been pumped to $|g\rangle$ at the start of the experiments.

Following the blink, we apply the blow-out pulse, repump the atoms and take an image. Then, the corrected blink survival probability $B$ can be extracted as:

$$B = \frac{P(I_1|I_0) + F_0 - 1}{SK(F_1 + F_0 - 1)(1 - R(1 - H))} - \frac{H(1 - K - L)}{K(1 - R(1 - H))}. \tag{S14}$$

We measure $B = 0.958(7)$ averaged over the array. This is the main source of error for detecting the ground state atoms. Implementation of a 698-nm laser system to coherently pump into $5s5p\,^3P_0$ would avoid heating the atoms by optical pumping and reduce this error.

**Rydberg state decay**   The main error in detecting the Rydberg state as losses comes from decay of Rydberg atoms to states that appear bright in the second image. For the $5s61s\,^3S_1$ state that is used in this work a characteristic timescale of $\tau_b = 168(14)\,\mu$s for decay to the $5s5p\,^3P_J$ is measured, with a branching ratio of about $6:3:1$ for $J = 2, 1, 0$ [5, 10]. The amount of time that atoms spend in the Rydberg state depends on the simulation, but also on the rate at which the Rydberg state is expelled by the tweezers afterwards. We characterize this rate by preparing atoms in $5s5p\,^3P_0$, applying a $\pi$-pulse to populate the Rydberg state, turning on the tweezers again and applying a second $\pi$-pulse after a variable wait duration. The second $\pi$-pulse is frequency-shifted to account for the AC Stark shift of the 813-nm light and its duration is first calibrated in a separate experiment. The fraction of atoms that successfully return to $5s5p\,^3P_0$ is measured to decay exponentially with a characteristic timescale of $\tau_e = 2.88(11)\,\mu$s. The decay of population of atoms in the Rydberg state is then described as:

$$P_r(t) \propto \begin{cases} e^{-t/\tau_b}, & t \leq t_r \\ e^{-t_r/\tau_b} e^{-(t - t_r)/\tau_c}, & t > t_r \end{cases}, \tag{S15}$$

where $\tau_c = \frac{\tau_b \tau_e}{\tau_b + \tau_e}$ is the combined decay lifetime and $t_r$ is the moment that the 813-nm light is turned on. With the maximal ramp duration of $7\,\mu$s and a $1.1\,\mu$s delay before turning on the 813-nm light, we assume $1.1\,\mu\text{s} \leq t_r \leq 8.1\,\mu$s. Solving for the population decayed to bright states at $t \to \infty$ this gives an estimation of $0.018 \leq X \leq 0.058$ for atoms in our experiments. Following the branching ratio of the decay and the repumping, we estimate that about $0.25X$ ends up in $5s5p\,^3P_0$.

It should be noted that this estimation neglects blackbody decay to other Rydberg states, which typically happens on a faster timescale than the decay to bright states [10, 11]. We believe this is justified because the decay couples mostly to nearby Rydberg states that have similar polarizabilities at 813 nm and have a similar lifetime. Furthermore, after such a decay event, the atom will most likely be in a $^3P_J$ series that does not directly couple to $5s5p\,^3P_0$. We do not consider the effects of such decays during the simulation.

## 4.2 Inference of the atom configuration from the measurement

Here, we will further place the experimental results in perspective by considering how experimental uncertainties affect the measured quantities. Our approach is based on two recent examples [12, 13]. For convenience, we will assume independent and equal errors for every site and do not consider correlated noise, although during prolonged Rydberg excitations, a single decay can trigger a cascade of errors [14].

Let us start by considering only a single atom, based on the individual errors listed and characterized in the previous subsection. We consider a simplified model including the optical pumping to $|\downarrow\rangle$, the Rydberg excitation, and a final detection step. The probability of detecting the atom as "1" and "0" in the final image can then be written as:

$$\begin{pmatrix} \text{``0''} \\ \text{``1''} \end{pmatrix} = \begin{pmatrix} 1 - \epsilon_p & \epsilon_n & 0 \\ \epsilon_p & 1 - \epsilon_n & 1 \end{pmatrix} \begin{pmatrix} P_g & 0 \\ P_r & 0 \\ 0 & 1 \end{pmatrix} \begin{pmatrix} p & 0 \\ 1 - p & 0 \end{pmatrix} \begin{pmatrix} 1 \\ 0 \end{pmatrix}. \tag{S16}$$

Because of post-selection, we assume a unity starting probability of having an atom present on the right-hand side. Then, there is a pumping step that prepares the atom in $|\downarrow\rangle$ with probability $p$ and loses the atom with probability $1 - p$. This is justified because a blow-out pulse effectively removes all atoms that are not pumped to $|\downarrow\rangle$. Following the pumping, there is a Rydberg excitation that leaves the pumped atoms in $|\downarrow\rangle$ with probability $P_g$ and in $|\uparrow\rangle$ with probability $P_r = 1 - P_g$. In the final detection stage, all the lost atoms are labeled as "1", while for the atoms that participated in the Rydberg excitation, we introduce an effective false positive probability $\epsilon_p$ of detecting an excitation when there was none and a converse false negative probability $\epsilon_n$.

We estimate a probability of pumping an atom into $|\downarrow\rangle$ of $p = SK \approx 0.989$ for our system. Likewise, $\epsilon_p$ and $\epsilon_n$ can be expressed as:

$$\epsilon_p = F_0 - B(F_1 + F_0 - 1)(1 - R(1 - H)), \tag{S17}$$

$$\epsilon_n = 1 - F_0 + \frac{BX}{20}(F_1 + F_0 - 1)(5 - 2R + H(15 + 2R)). \tag{S18}$$

Substituting the values from Section 4.1, we estimate $\epsilon_p \approx 4.8\,\%$ and $\epsilon_n \approx 1\,\%$ for the experiments used here.

Equation (S16) can be simplified to:

$$\begin{pmatrix} "0" \\ "1" \end{pmatrix} = p \begin{pmatrix} 1 - \epsilon_p & \epsilon_n \\ \epsilon_p & 1 - \epsilon_n \end{pmatrix} \begin{pmatrix} P_g \\ P_r \end{pmatrix} + (1 - p) \begin{pmatrix} 1 \\ 0 \end{pmatrix}, \tag{S19}$$

which is more compactly written as:

$$\vec{P}_b = p\mathbf{M}\vec{P}_\psi + (1-p)\vec{O}, \tag{S20}$$

where $\vec{P}_b$ is a probability distribution of measured outcomes, $\mathbf{M}$ is a detection matrix for atoms that participated in the simulation, $\vec{P}_\psi$ is a probability distribution of the state that the atom was in following the Rydberg excitation and $\vec{O}$ is an offset vector that takes into account the fraction of atoms that was not pumped to $|\downarrow\rangle$ and thus never participated in the simulation. The general goal is to infer a distribution $\vec{P}_\psi$ based on the measured outcomes $\vec{P}_b$.

If we assume equal and independent errors for all atoms, we can now extend Equation (S20) to $N$ atoms as:

$$\vec{P}_b^N = p^N \mathbf{M}^N \vec{P}_\psi^N + (1 - p^N)\vec{O}^N. \tag{S21}$$

$\vec{P}_b^N$ now has the measured probabilities for bitstrings "00...00" through "11...11", $\vec{P}_\psi^N$ describes the inferred probabilities of product states $|\downarrow\downarrow\ldots\downarrow\downarrow\rangle$ through $|\uparrow\uparrow\ldots\uparrow\uparrow\rangle$, and $\mathbf{M}^N = \mathbf{M} \otimes \cdots \otimes \mathbf{M}$. The offset $\vec{O}^N$ also is a probability distribution of bitstrings, but its distribution is not trivially found. For the single atom case, we could simply assume that every atom that was not pumped would be detected as "1", but for more atoms this can not be assumed. There is no information on what other atoms that have been pumped to $|\downarrow\rangle$ do after the Rydberg excitation if a neighboring atom wasn't pumped, so we cannot use our model to directly link $\vec{O}^N$ to experimentally observed values. A consequence of this is that it is impossible to use the inverse of $\mathbf{M}^N$ to solve for $\vec{P}_\psi^N$.

A more controlled way of estimating $\vec{P}_\psi^N$ is by optimizing the individual probabilities inside it while minimizing a cost function [12]. We infer distributions $\vec{P}_\psi^N$ and $\vec{O}^N$ using the SciPy implementation of the SLSQP algorithm [15] with scalar L2 cost function

$$C = \left\| p^N \mathbf{M}^N \vec{P}_\psi^N + (1 - p^N)\vec{O}^N - \vec{P}_b^N \right\|, \tag{S22}$$

and boundary conditions $\sum_{i=1}^{2^N} \vec{P}_{\psi,i}^N = 1$ and $\sum_{i=1}^{2^N} \vec{O}_i^N = 1$ to ensure a physically valid solution.

# 5  Supplementary Notes: Influence of trap deformations

To further elaborate on the trap deformation mentioned in the Methods, we present measurements showing the influence of different effects on the magnetization out come in the $\langle n_\uparrow \rangle = 4$ state. It should be noted that this is not an exhaustive study and rather serves as a potential explanation of the observed asymmetry.

## 5.1  Trap deformations due to interference

To calculate holograms using the WGS algorithm, we define a target intensity pattern in a computational grid that has the same pixel number as the SLM chip. For tightly spaced patterns, the distance between neighboring tweezers is comparable to the beam diameter and neighboring tweezers can interfere. As the WGS algorithm only considers the intensity distribution at the coordinates of the tweezers, it cannot correct for interference in the current implementation of the algorithm. It should be noted that a gradient-based optimization algorithm that takes into account the finite size of the tweezers shows great promise in mitigating these types of errors [16].

Although we do not control the final phase distribution of the WGS algorithm, we can provide different initial tweezer phases, which result in different final holograms. We compare the results of performing the same sweep to the $\langle n_\uparrow \rangle = 4$ state for two holograms. The first hologram had random initial phases in the calculation. Supplementary Fig. 5a shows the projected optical phases, calculated with a fast Fourier transform (FFT) of the hologram. In line with the random initial phases, the optical phases are not ordered. Supplementary Fig. 5b shows the measured magnetization, which shows a preference in exciting region-of-interest (ROI) 3. The second hologram had ordered initial phases, in which neighboring tweezers had a phase difference of $\pi$. The resulting hologram after the WGS calculation still had close to these values, with the exception of ROIs 4 and 5 for which the phase had increased with roughly $\pi$, see Supplementary Fig. 5c. This highlights the ability of the WGS algorithm to produce the same intensity distribution with different optical phase distributions. In Supplementary Fig. 5d, the magnetization measured using this second hologram is plotted. The excitation of ROI 4 is favored, displaying the sensitivity of the experiment to the exact hologram used.

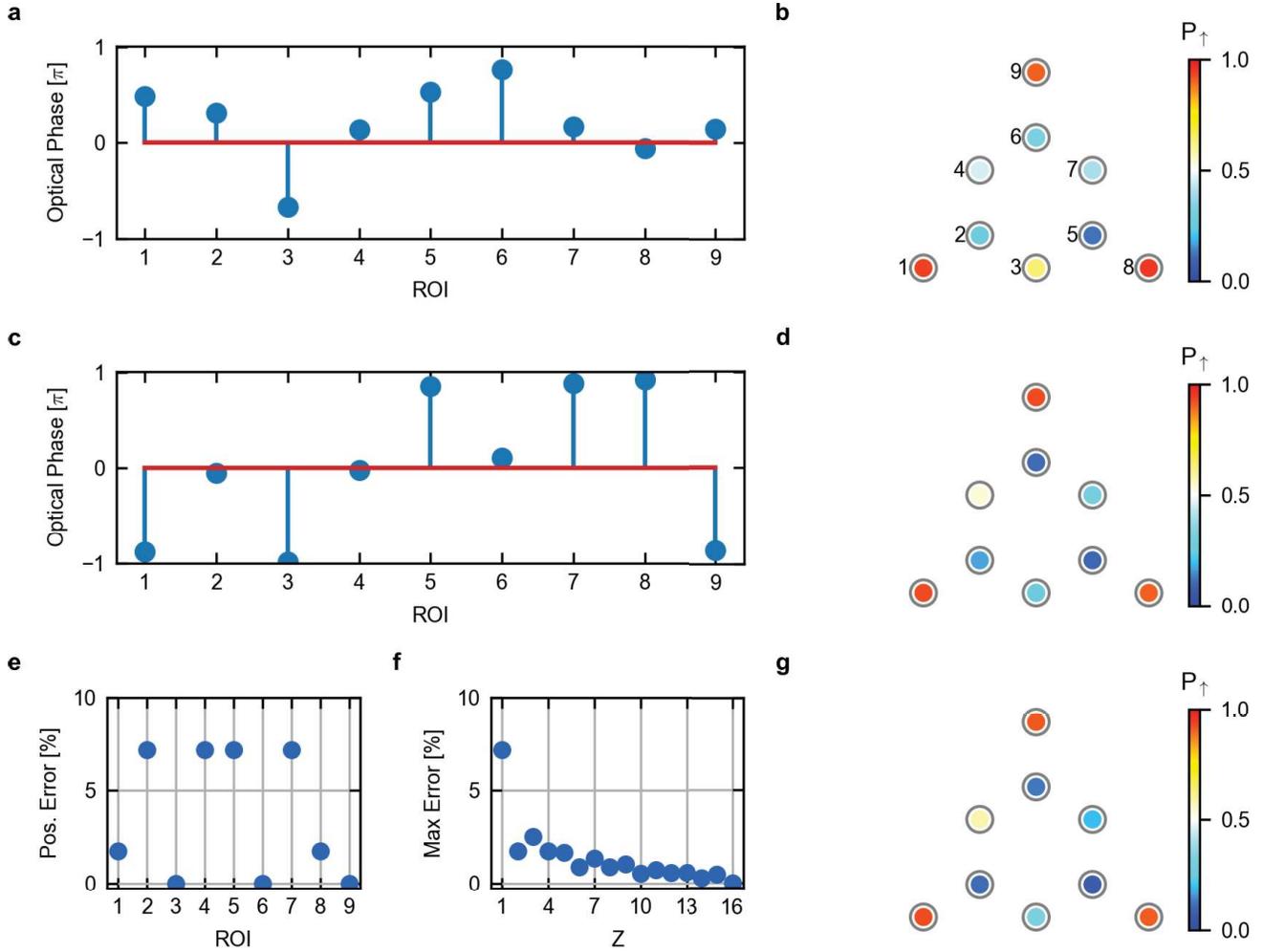

**Supplementary Figure** 5: **a,** Optical phases per ROI as computed with an FFT for the hologram with random initial optical phases. **b,** The magnetization when preparing an $\langle n_\uparrow \rangle = 4$ state, using the same scan parameters as reported in Extended Data Tab.1, using the pattern with optical phases displayed in **a**. **c,** The optical phases per ROI after providing an ordered set of initial phases, where each neighboring trap had a phase difference of $\pi$. Except for ROI 4 and 5, this relation is roughly maintained after calculation with the WGS algorithm. **d,** The measured magnetization with the same scan as **b**, for this hologram shows that ROI 4 is now preferred instead of ROI 3. **e,** The relative position error per ROI at $Z = 1$. **f,** Maximal relative position error per ROI from rounding a hexagonal pattern on a square grid versus the factor Z with which the resolution is increased in calculation (see text). **g,** The observed magnetization for a Z=4 hologram, for which we observe that the preference has shifted to ROI 4. Furthermore, ROIs 2, 5 and 6 have become more uniform. This can be compared to the random initial phases and Z=1 pattern presented in **b**.

The difference in magnetization is not explained by a different distribution of trap depths. Since there is only a single microscope objective installed on the experiment, there is no direct way of observing the tweezer pattern that is projected on the atoms. The best feedback on the trap depth therefore comes from atoms inside the trap. By performing spectroscopy on the $5s^2\,{}^1S_0 \leftrightarrow 5s5p\,{}^3P_1$ ($m_J = \pm 1$) transition, we measure the differential AC Stark shift — proportional to the trap depth — of each tweezer. For each tweezer pattern, we minimize the spread of the trap depths by iteratively calculating holograms where the measured trap depth distribution of the previous hologram is the input for the next target intensity distribution [17]. Typically, this results in a standard deviation of the trap depths of less than $1\,\%$. We note that this process successfully creates equal trap depths, but does not guarantee every trap geometry is the same. For such an optimization, the trap frequencies could be used as an optimization parameter, which has not been done here [18].

## 5.2 Corrections from designing a hexagonal pattern on a square grid

The smallest distance that can be calculated with a discrete Fourier transform is given as:

$$d_{\text{FFT}} = \frac{\lambda f}{mL}, \tag{S23}$$

where $\lambda$ is the wavelength of the tweezer light, $f$ is the effective focal length of the objective and $mL$ is the demagnified size of SLM. Intuitively, one can see the meaning of this Fourier unit $d_{\text{FFT}}$ as the translation in the focal plane when applying a phase gradient $\phi_n = n2\pi/N$ on the SLM, where $n$ is the pixel number and $N$ the total number of pixels in the horizontal or vertical direction. In our system $d_{\text{FFT}} \approx 0.45\,\mu\text{m}$ in the focal plane of the objective. For the strongest interactions used in this work, a separation of $d = 7\,d_{\text{FFT}}$ is used, corresponding to $R \approx 3.15\,\mu\text{m}$.

The hexagonal lattice underlying the investigated fractal has coordinates that are multiples of $\sqrt{3}$ when converted to Cartesian coordinates. This results in non-integer coordinates of the target tweezer positions on the square grid of the SLM. Since the WGS algorithm can only calculate integer coordinates, this leads to rounding errors that degrade the threefold symmetry of the pattern. In Supplementary Fig. 5e, the relative error from rounding is given for $d = 7\,d_{\text{FFT}}$. The maximal error is $7.2\,\%$ for ROIs 2, 4, 5, 7. Due to the $R^{-6}$ scaling of the interaction, errors of $7.2\,\%$ amount to a $50\,\%$ change in the interaction strength. One way to compensate for this effect is by increasing the computational space used in the calculation of the holograms [16, 19]. This gives an increased resolution of a factor $Z$ at the cost of increasing the number of pixels by $Z \times Z$. In the plane of the atoms, every Fourier unit becomes $Z$ times smaller and as a consequence, rounding errors become less relevant. It should be noted that the resulting hologram is also $Z \times Z$ times bigger. We crop the center $1024 \times 1024$ pixels at the end of the calculation and display only those on the SLM. Effectively, this results in holograms that contain smaller gradients down to $2\pi/NZ$, with $N$ the number of pixels per side of the SLM chip. The expected decrease in maximal relative rounding error for factors up to $Z = 16$ is presented in Supplementary Fig. 5f. The error decreases for increasing $Z$ and is almost zero at $Z = 16$.

For tests on atoms, we compare the $Z = 1$ and $Z = 4$ holograms. Choosing $Z$ such that the total pixel size is a power of 2 is favorable, because fast Fourier transforms work best at dimensions $2^m \times 2^m$. To reduce computational time when optimizing the trap depth uniformity of the pattern, we favored $Z = 4$ over $Z = 16$ in this case. In Supplementary Fig. 5g, we present the average magnetization of the atoms in the tweezers formed by the $Z = 4$ hologram. Both holograms had similar uniformities of trap depth and survival of the experimental sequence without UV light. The $Z = 1$ pattern (Supplementary Fig. 5b) favors excitations of ROI 3, whereas the $Z = 4$ pattern favors excitation of ROI 4. This highlights the sensitivity of the measurement to small displacements.

We did not see a further improvement when using both zeropadding and an initial guess of the optical tweezer phases. For the data in the main text, we used the $Z = 1$ pattern with the initial guess of alternating phases as presented at the start of this subsection, because it gave a slightly more uniform result.